\documentclass[prb,amsmath,twocolumn,letterpaper]{revtex4-2}
\usepackage{graphicx}
\usepackage{amssymb}
\begin{document}

\title{{Absence of Luther-Emery Superconducting Phase in the Three-Band Model for Cuprate Ladders}}
\author{Jeong-Pil Song}
\affiliation{Department of Physics, The University of Arizona Tucson, AZ 85721}
\author{Sumit Mazumdar}
\affiliation{Department of Physics, The University of Arizona Tucson, AZ 85721}
\author{R. Torsten Clay}
\affiliation{Department of Physics \& Astronomy, and HPC$^2$ Center for Computational Sciences, Mississippi State University, Mississippi State, 
 MS 39762}
 \date{today}
\begin{abstract}
  Correlated-electron theories of superconductivity in layered
  cuprates often start from the premise of a gapped spin-liquid phase
  proximate to the superconducting state.  This assumption is
  justified based on analytical and numerical demonstrations of a
  superconducting Luther-Emery phase in the doped 2-leg one-band
  Hubbard ladder, and the perceived analogy between coupled ladders
  and the two dimensional CuO$_2$ layer.  We demonstrate from accurate
  density matrix renormalization group studies the absence of the
  superconducting Luther-Emery phase in the doped 2-leg three-band
  ladder consisting of both copper and oxygen, even as the spin gap is
  large in the undoped three-band ladder. For realistic oxygen-oxygen
  hopping and Hubbard repulsion on the oxygen atoms, density-density
  rather than pairing correlations are dominant at long range.  This
  result is equally valid whether or not the oxygens outside the
  ladder proper, over and above the rung and leg oxygens, are included
  in the computation.  These results demonstrate the critical
  importance of oxygen orbitals, and raise disturbing questions about
  the applicability of many of the existing correlated-electron
  theories of superconductivity.
\end{abstract}
\maketitle

\section{Introduction}
More than three decades after the discovery of high temperature
superconductivity (SC) in cuprates, there is no consensus on the
mechanism of the phenomenon.  There is broad agreement that the
undoped parent antiferromagnetic compounds can be described within the
Cu-only one-band two-dimensional (2D) Hubbard model, which ignores the
O-ions entirely.  Proximity of SC to antiferromagnetism has led to the
widely held belief that the mechanism of SC can also be found within
an effective weakly-doped single-band Hubbard model
\cite{Anderson87b,Anderson04a,Lee06b,Scalapino12a} based on the claim
that the spins on the Cu-sites and on the dopant-induced holes on the
O-sites form local spin-singlets that behave like double occupancies
in the single-band Hubbard model \cite{Zhang88a}.  The list of
approximate correlated-electron theories that find SC within the
weakly doped one-band Hubbard model is long, but accurate numerical
studies have consistently found that superconducting correlations are
suppressed by the Hubbard $U$ for carrier concentrations believed to
be appropriate for the superconductors \cite{Zhang97b,Aimi07a}.
Recent very careful study using two distinct and complementary
numerical approaches to calculating superconducting pair-pair
correlations has concluded that SC is absent in the square lattice
Hubbard model proximate to $\frac{1}{2}$-filling \cite{Qin20a}.
Inclusion of second neighbor hopping $t^\prime$ beyond nearest
neighbor (n.n.) does not change this conclusion
\cite{Huang01a,Chung20a}. Signatures of pair-correlations enhanced by
the Hubbard $U$ have been found uniquely at $\frac{1}{4}$-filling
\cite{Gomes16a}, far from the carrier concentration believed to be
appropriate for the cuprates. A recent density matrix renormalization
group (DMRG) calculation has claimed transition from $p$-wave to
$d$-wave SC within the hole-doped triangular lattice Hubbard model,
but once again at fillings very far away from $\frac{1}{2}$
\cite{Venderley19a}.
\begin{figure}[b]
  \centerline{\raisebox{0.2in}{\resizebox{1.45in}{!}{\includegraphics{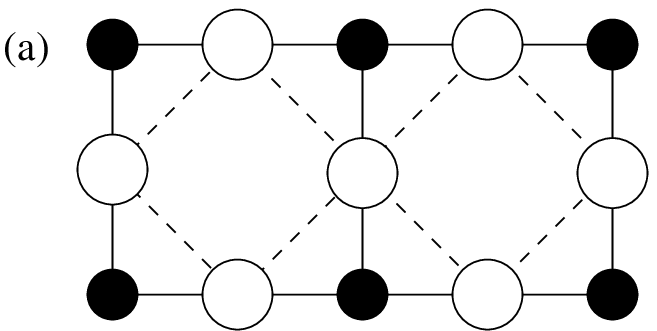}}}\hspace{0.1in}\resizebox{1.45in}{!}{\includegraphics{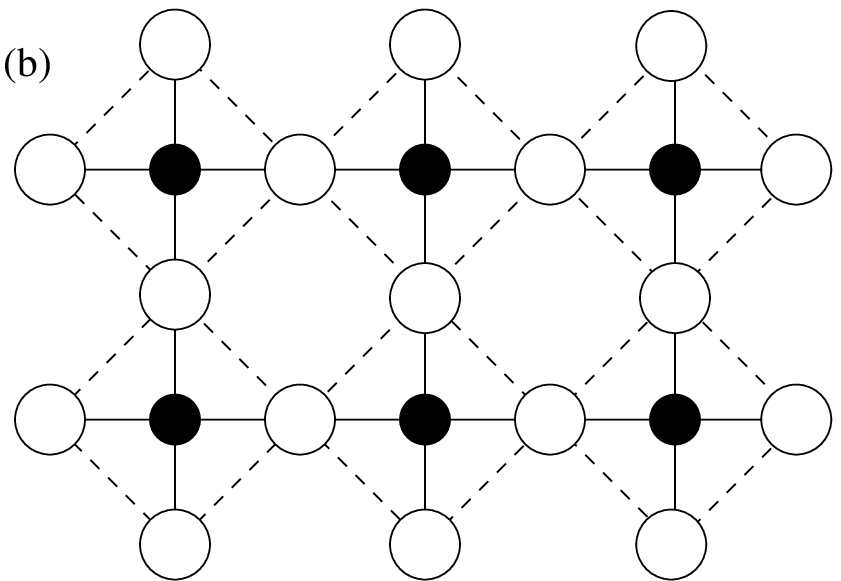}}}
  \caption{Cuprate ladder geometries considered by us. Filled (open)
    circles represent copper (oxygen) atoms. (a) Ladder with three
    oxygen atoms per rung. (b) Ladder with five oxygen atoms per
    rung.}
  \label{fig-lattice}
\end{figure}

A key reason for the continued application of the one-band Hubbard
model to cuprates is the repeated finding that the ground state of the
weakly doped 2-leg one-band Hubbard ladder is a Luther-Emery liquid,
with gapless charge and gapped spin modes
\cite{Luther74a,Noack94a,Noack97a,Balents96a,Hur09a,Dolfi15a,Gannot20a}.
Such a spin-gap proximity effect has been considered essential for SC
within an entire class of theories
\cite{Anderson87b,Lee06b,Emery97a,Hur09a}.  DMRG calculations, highly
precise for one dimensional (1D) Hamiltonians, find slower than $1/r$
decay of the superconducting pair-pair correlation $P(r)$ in the doped
2-leg one-band Hubbard ladder, where $r$ is the interpair separation
\cite{Noack94a,Noack97a,Dolfi15a,Gannot20a}. Power law decay slower
than $1/r$ is a requirement as well as signature of quasi-long
range order in one dimension \cite{Feiguin08a}. Strong superconducting correlations in
the doped ladder is a consequence of spin-singlet formation on the
undoped ladder rungs \cite{Dagotto96a}.  The DMRG results therefore
have lent credence to the viewpoint that some variant of the 2D
one-band Hubbard model with minor modifications might still yield SC.
\begin{figure}[b]
  \centerline{\resizebox{3.0in}{!}{\includegraphics{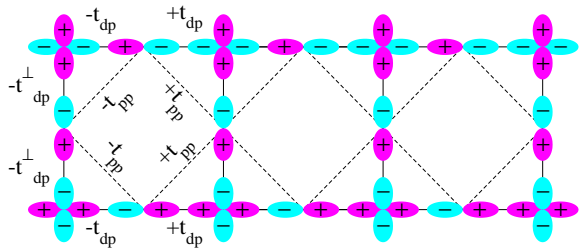}}}
  \caption{(Color online) Orbital parity and sign convention for the
    hopping matrix elements for the ladder geometry of
    Fig.~\ref{fig-lattice}(a). Orbital parity and signs for the
    lattice of Fig.~\ref{fig-lattice}(b) follow a similar convention
    (see text).}
  \label{fig-signs}
\end{figure}

A realistic description of the cuprate ladder, however, should include
the oxygen(O)-ions (see Fig.~\ref{fig-lattice}).  Surprisingly, few
authors have investigated the appropriateness of replacing the more
complete three-band 2-leg ladder Hamiltonian that includes the O-ions
with the one-band Hubbard Hamiltonian
\cite{Jeckelmann98a,Nishimoto02a}.  We have performed accurate DMRG
calculations on long three-band 2-leg ladders for parameters
appropriate for real cuprates.  In order to reach the longest ladders
possible, we have performed the bulk of our calculations for the
geometry of Fig.~\ref{fig-lattice}(a), with three O-ions per ladder
rung. We have also performed limited calculations of superconducting
pair correlations for a shorter ladder with the geometry of
Fig.~\ref{fig-lattice}(b), with O-ions outside the ladder bonded to
the Cu sites \cite{Nishimoto02a}.  We find that that the
distance-dependence of the superconducting pair correlations are
independent of geometry, and are strongly suppressed for both
geometries of Fig.~\ref{fig-lattice} at the doping traditionally
assumed to correspond to the maximum of the superconducting dome
within a one-band picture, $\delta\sim0.125$.  We further determine
the Luther-Emery correlation exponent, $\kappa_\rho$, in the
thermodynamic limit through fits of the charge density, finding
$\kappa_\rho<1$ for $\delta=0.125$, consistent with the suppression of
pairing.  We present physical arguments within an effective
Hamiltonian that explain the suppression of the superconducting pair
correlations and their rapid decay.  These results have profound
implications for any realistic modeling of cuprates.

\section{Theoretical model, parameters and computational techniques}

The one-band ladder Hamiltonian has parameters $U$, the Hubbard
repulsion; and the leg and rung hopping integrals $t$ and $t_{\perp}$,
respectively.  It is customary to express $U$ and $|t_{\perp}|$ in
units of $|t|$.  Here we consider the three-band ladder Hamiltonian,
\begin{align}
&H = \Delta_{\rm dp}\sum_{i\sigma} p^\dagger_{i,\sigma}p_{i,\sigma}
     +\sum_{\langle ij \rangle, \lambda, \sigma}t_{\rm dp}^{{\perp},i,j}
     (d^\dagger_{i,\lambda,\sigma}p_{j,\sigma}+H.c.)\nonumber \\
&+\sum_{\langle ij \rangle,\lambda, \sigma}t_{\rm dp}^{i,j}(d^{\dagger,i,j}_{i,\lambda,\sigma}p_{j,\sigma}+H.c.)
     +\sum_{\langle ij \rangle, \sigma}t_{\rm pp}^{i,j} 
     (p^\dagger_{i,\sigma}p_{j,\sigma}+H.c.)\nonumber \\
 &+U_{\rm d}\sum_{i,\lambda} d^\dagger_{i,\lambda,\uparrow}d_{i,\lambda,\uparrow}d^\dagger_{i,\lambda,\downarrow}d_{i,\lambda,\downarrow}
     +U_{\rm p}\sum_j p^\dagger_{j,\uparrow}p_{j,\uparrow}p^\dagger_{j,\downarrow}p_{j,\downarrow}
     \label{hamiltonian}
\end{align}

Here $d^\dagger_{i,\lambda,\sigma}$ creates a hole with spin $\sigma$
on the $i$th Cu $d_{x^2-y^2}$ orbital on the $\lambda$-th leg
($\lambda=1,2$), $p^\dagger_{j,\sigma}$ creates a hole on the
n.n. rung oxygen O$_{\rm R}$ or leg oxygen O$_{\rm L}$; $t_{\rm
  dp}^{\perp,i,j}$ and $t_{\rm dp}^{i,j}$ are n.n. ladder rung and leg
Cu-O hopping integrals and $t_{\rm pp}^{i,j}$ is the n.n. O-O hopping
integral.  For the lattice of Fig.~\ref{fig-lattice}(a), the Cu-O
hopping matrix elements along the legs of the ladder have the sign
convention $t_{\rm dp}^{i,j} = -t_{\rm dp}$ for $j =i+{\hat x}/2$ and
= $t_{\rm dp}$ for $j =i-{\hat x}/2$. Along the rungs of the ladder
$t^{\perp,i,j}_{\rm dp}=-t^\perp_{\rm dp}$.  Similarly $t_{\rm
  pp}^{i,j}={\pm} t_{\rm pp}$, with the plus and minus signs occurring
for $j=i+{\hat x}/2\pm{\hat y}/2$ and $j=i-{\hat x}/2\pm{\hat y}/2$,
respectively (see Fig.~\ref{fig-signs}).  For the lattice of
Fig.~\ref{fig-lattice}(b) we choose the orbital parities so that the
added $t_{\rm dp}^{\perp,i,j}$ bonds are $-t^\perp_{\rm dp}$ and the
added $t_{\rm pp}^{i,j}$ are positive and negative for $j=i-{\hat
  x}/2\pm{\hat y}/2$ and $j=i+{\hat x}/2\pm{\hat y}/2$, respectively.

$U_{\rm d}$ and $U_{\rm p}$ are the onsite repulsions on the Cu and O
sites, and $\Delta_{\rm dp} = \epsilon_p - \epsilon_d$ where
$\epsilon_p$ ($\epsilon_d$) is the site energy of O (Cu). In what
follows all energies are in units of $|t_{\rm dp}|$.  Most of our
calculations are for $t_{\rm dp}=t_{\rm dp}^{\perp}=1$, $U_{\rm d}=8$,
$\Delta_{\rm dp}=3$; we show a few results also for $t_{\rm
  dp}^{\perp}>1$.  We show results for $t_{\rm pp}=0$ and 0.5, for two
different $U_{\rm p}=0, 3$; the parameter set $t_{\rm pp}=0.5, U_{\rm
  p}=3$ is nearly identical to those used previously
\cite{Jeckelmann98a,Nishimoto02a}. We also show results of
calculations of superconducting pair correlations for the parameters
$t_{\rm pp}=0.6, U_{\rm p}=4$, as these are very close to those
derived from recent first principles calculations \cite{Hirayama18a}.
Our calculations are performed for different dopings $\delta$, where
$1+\delta$ is the average hole concentration per Cu-ion ($\delta=0$
for the undoped ladder).
\begin{figure}
  \centerline{\resizebox{3.4in}{!}{\includegraphics{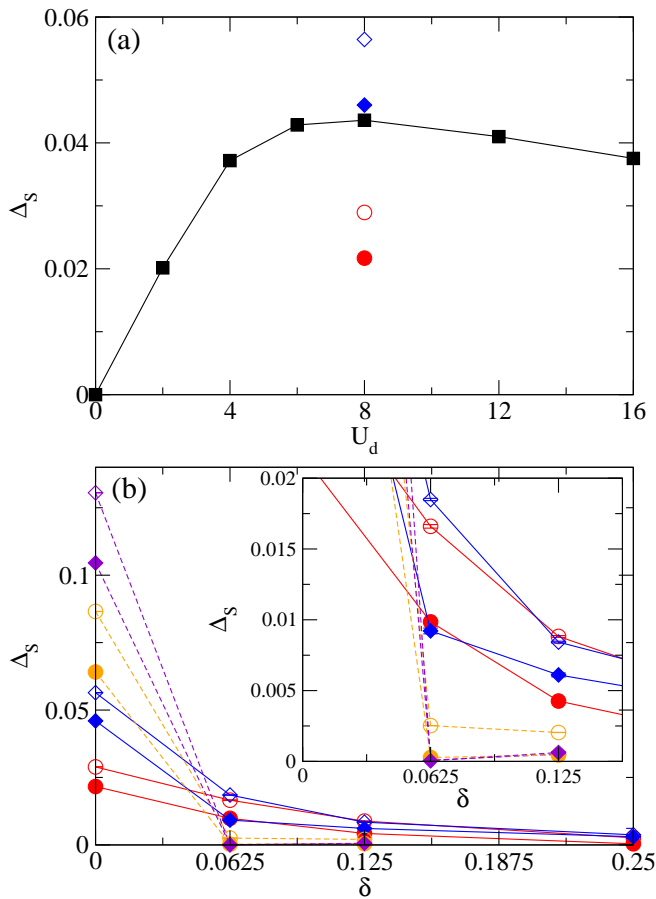}}}
  \caption{(Color online) (a) Spin gap versus $U_{\rm d}$ in the
    undoped three-band ladder.  Squares: $U_{\rm p}$=$U_{\rm d}/2$,
    $t_{\rm pp}$=0.5; Open circle: $U_{\rm p}$=0, $t_{\rm pp}$=0; Open
    diamond: $U_{\rm p}$=0, $t_{\rm pp}$=0.5; Filled circle: $U_{\rm
      p}$=3, $t_{\rm pp}$=0; Filled diamond: $U_{\rm p}$=3, $t_{\rm
      pp}$=0.5.  In all cases $t_{\rm dp}^{\perp}=1$.  (b) Same as a
    function of doping $\delta$. Solid (dotted) lines correspond to
    $t_{\rm dp}^{\perp}=1$ ($|t_{\rm dp}^{\perp}|=1.25$). Open
    circles: $U_{\rm p}$=0, $t_{\rm pp}$=0; Open diamonds: $U_{\rm
      p}$=0, $t_{\rm pp}$=0.5; Filled circles: $U_{\rm p}$=3, $t_{\rm
      pp}$=0.0; Filled diamonds: $U_{\rm p}$=3, $t_{\rm pp}$=0.5;
    Lines are guides to the eye. The inset magnifies the small doping
    region. DMRG truncation and finite-size extrapolation errors are
    smaller than the symbol size.}
  \label{fig-gaps}
\end{figure}

Our DMRG calculations are done with open boundary conditions, with
Cu-O-Cu rungs at both ends (see Fig.~\ref{fig-lattice}(a)).
Calculations used the ITensor library \cite{itensor} with a two-site
DMRG update, particle number and $S_z$ conservation, and real-space
parallelization \cite{Stoudenmire13a}.  For the geometry of
Fig.~\ref{fig-lattice}(a), we have considered ladders with length $L$
up to $L$=40 rungs (198 sites) for the undoped case, and with up to
$L=96$ (478 sites) for doped cases, with bond dimension $m$ up to
19,000.  The minimum DMRG truncation error was of order 10$^{-8}$; we
extrapolated energies and correlation functions to zero truncation
error as detailed in \cite{Supplemental}.  For the geometry of
Fig.~\ref{fig-lattice}(b), our calculations of superconducting pair
correlations (see below) are for $L=64$ (450 sites).  These
calculations are for the longest ladders containing both Cu and O with
the largest $m$ to date.

\section{Computational results} 

In what follows, the computational results are for the geometry of Fig.~\ref{fig-lattice}(a), unless
it is explicitly mentioned that they are for the geometry of Fig.~\ref{fig-lattice}(b).

\subsection{Charge densities}

In Table I we have given the calculated $L \to \infty$ extrapolated
charges (see Supplemental Material \cite{Supplemental} for the details
of the extrapolation procedure) on the Cu-ions $\langle n_{Cu}
\rangle$, rung O-ions $\langle n_{O_{\rm R}} \rangle$ and leg O-ions
$\langle n_{O_{\rm L}} \rangle$, respectively, for $\delta=0$ and
$\delta=0.125$, which are representative for other $\delta$ (see
Supplemental Material \cite{Supplemental}).  The doping-induced
increase in $\langle n_{Cu} \rangle$ is very small, with the bulk of
the doped charge going to O-ions.  Importantly, given that there occur
two leg O-ions corresponding to each rung O, the overall increase in
population due to doping is larger for O$_L$ than O$_R$.  Our
calculated charge-densities are close to those obtained previously for
shorter three-band ladders for similar parameters
\cite{Nishimoto02a}. The calculated Cu-ion charge densities are very
close to those in reference \onlinecite{Nishimoto02a}, where however,
the calculations are for the geometry of
Fig.~\ref{fig-lattice}(b). The nearly same Cu-ion charge densities for
the two geometries is because the hole density on the outer O-ions are
quite small \cite{Nishimoto02a}. This already suggests similar
behavior of superconducting pair correlations in the two geometries,
which we indeed find to be true.
\begin{table}[t]
\caption{Average extrapolated charge densities on Cu and O-sites in
  the undoped and doped ($\delta=0.125$) three-band Hubbard
  Hamiltonian for $U_{\rm d}=8$.}  \centering
\begin{ruledtabular}
\begin{tabular}{cccccccc}
$U_{\rm p}$  & $t_{\rm pp}$ &
\multicolumn{2}{c}{$\langle n_{\rm Cu}\rangle$} &
\multicolumn{2}{c}{$\langle n_{O_{\rm R}}\rangle$} &
\multicolumn{2}{c}{$\langle n_{O_{\rm L}}\rangle$} \\
& &
$\delta$=0 &
$\delta$=0.125 &
$\delta$=0 &
$\delta$=0.125 &
$\delta$=0 &
$\delta$=0.125 \\
\hline
0 & 0.0 & $0.81$ & $0.82$ & $0.13$ & $0.19$ & $0.12$ & $0.20$ \\
0 & 0.5 & $0.73$ & $0.75$ & $0.22$ & $0.28$ & $0.15$ & $0.22$ \\
3 & 0.0 & $0.82$ & $0.84$ & $0.12$ & $0.18$ & $0.12$ & $0.19$ \\
3 & 0.5 & $0.75$ & $0.78$ & $0.20$ & $0.26$ & $0.14$ & $0.21$ \\
\end{tabular}
\end{ruledtabular}
\end{table}
 
\subsection{Spin gaps, doped versus undoped}

Fig.~\ref{fig-gaps}(a) shows the extrapolated spin gaps $\Delta_{\rm
  S}=E(S_z=1)-E(S_z=0)$, where $E(...)$ is the lowest energy of the
state with total $z$-component of spin $S_z$, for $\delta=0$, $t_{\rm
  dp}^{\perp}=1$ and for a range of $U_{\rm d}$ for $U_{\rm p} =
U_{\rm d}/2$ (see Supplemental Material \cite{Supplemental} for
details of the extrapolation procedure).  For parameters for which
previous calculations exist in the literature (for e.g., $U_{\rm
  d}=8$, $t_{\rm dp}^{\perp}=1$, $t_{\rm pp}=U_{\rm p}=0$) our
calculated $\Delta_{\rm S}$ are the same as before
\cite{Jeckelmann98a}.  The increase in $\Delta_{\rm S}$ with $t_{\rm
  pp}$ for $\delta=0$ has been noted before \cite{Nishimoto02a}, and
can be understood within perturbation theory \cite{Eskes93a}.  Nonzero
$U_{\rm p}$ suppresses $\Delta_{\rm S}$ strongly. The behavior of
$\Delta_{\rm S}$ versus $U_{\rm d}/|t_{\rm pd}|$ is very similar to
that versus $U/|t|$ in one-band ladders, where also a maximum near
$U/|t|=8$ is observed \cite{Noack94a}. Undoped three- and one-band
models are thus indeed similar.

In Fig.~\ref{fig-gaps}(b) we have shown the doping dependence of the
extrapolated spin gaps (see Supplemental Material \cite{Supplemental}
for details of the extrapolations). We have included additional data
points for $t_{\rm dp}^{\perp}=1.25$ here.  $\Delta_{\rm S}$ is
suppressed strongly with doping, as is also true within the one-band
ladder. Importantly, it appears that the larger is the spin gap in the
undoped state, the more rapid is the suppression of the spin gap with
doping. This is particularly obvious from comparison of $t_{\rm
  dp}^{\perp}=1.25$ versus 1.0.  Equally interesting is the strong
enhancement of $\Delta_{\rm S}$ in the undoped state when $t_{\rm pp}
\neq 0$, as noted above, but very rapid suppression of the same upon
doping.  We further note that the detrimental effects of nonzero
$U_{\rm p}$ and $t_{\rm pp}$ are synergistic, as seen from the inset
of Fig.~\ref{fig-gaps}(b), where the spin gap at $\delta=0.0625$ for
($U_{\rm p}$, $t_{\rm pp}$) = (3.0, 0.0) and (3.0, 0.5) are the same,
even as for $\delta=0$ the spin gap is larger by more than a factor of
2 for ($U_{\rm p}$, $t_{\rm pp}$ = 3.0, 0.5).  The very large
$\Delta_{\rm S}$ for $\delta=0$, $t_{\rm dp}^{\perp}=1.25$, along with
$\Delta_{\rm S} \simeq 0$ for the doped cases here are in agreement
with previous one-band calculation of the spin gap in the doped ladder
\cite{Noack96a} for $U=8$ and $t_{\perp}/t>1.5$.

The very rapid diminishing of $\Delta_{\rm S}$ with doping already
suggests the suppression of the superconducting correlations we find
(see below).
\begin{figure}
  \centerline{\resizebox{3.0in}{!}{\includegraphics{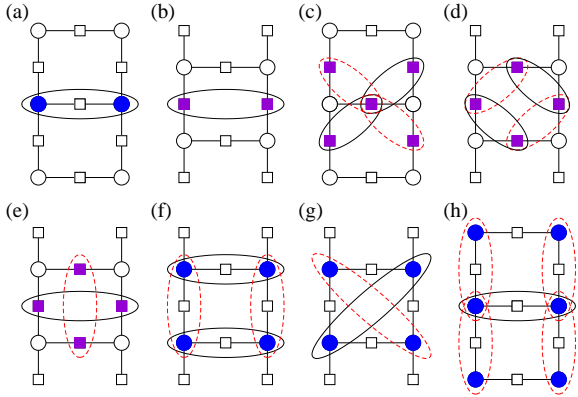}}}
  \caption{(Color online) Pair symmetries we considered. Ellipses
    represent singlets (a) copper rung singlet, (b) oxygen leg
    singlet, (c) type 0, (d) type 1, (e) type 2, (f) type 3, (g) type
    4, and (h) type 5. In (d)-(h) singlets represented by solid and
    dashed ellipses have opposite sign; for pair types (c)-(h) we
    considered both ``s-wave'' pairing, with all singlets having
    positive sign, and ``d-wave'' pairing, with the solid (dashed)
    singlets having sign $+1$ ($-1$).}
\label{fig-pairs}
\end{figure}

\subsection{Superconducting pair-pair correlations}
\label{pr}

\subsubsection{Rung singlet pairs}

\begin{figure}
  \centerline{\resizebox{3.4in}{!}{\includegraphics{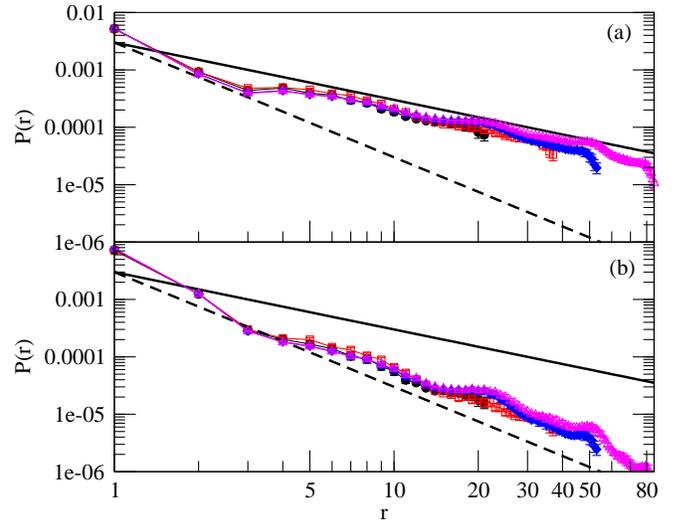}}}
  \caption{(Color online) Pair-pair correlation function $P(r)$ with
    $U_{\rm d}$ = 8 and doping $\delta=0.0625$ as a function of the
    rung-rung distance $r$, for (a) $U_{\rm p}=0$ and $t_{\rm pp}=0$,
    and (b) $U_{\rm p}=3$ and $t_{\rm pp}=0.5$. Circles, squares,
    diamonds, and triangles are for 32, 48, 64, and 96 rung ladders,
    respectively.  $P(r)$ data is extrapolated in the DMRG truncation
    error; error bars are calculated from lattice averaging (see
    text). The solid (dashed) lines are power laws $r^{-1}$
    ($r^{-2}$).}
  \label{fig-length}
\end{figure}

\begin{figure}
  \centerline{\resizebox{3.4in}{!}{\includegraphics{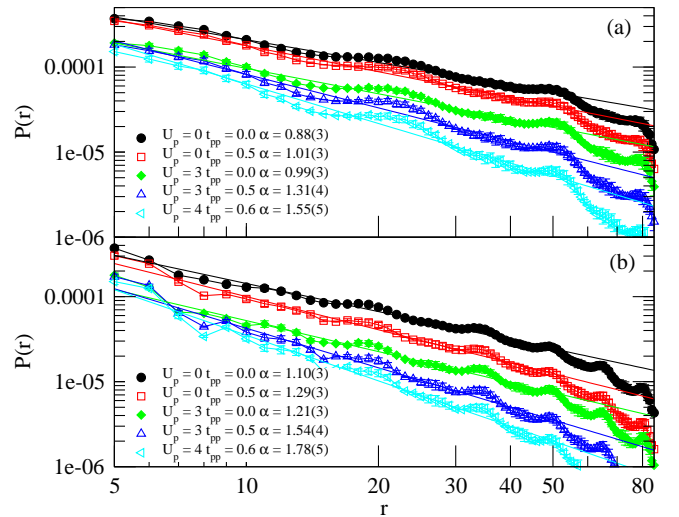}}}
  \caption{(Color online) $P(r)$ as a function of the rung-rung
    distance $r$ for a 96-rung ladder with $U_{\rm d}$ = 8, for (a)
    doping $\delta$ = 0.0625, and (b) $\delta$ = 0.1250.  Lines are
    linear fits of $P(r)$ for 10 $\leq r \leq$ 48; the power law
    exponents $\alpha$ are given in the figure legends.}
  \label{fig-96rung}
\end{figure}

For direct comparison of superconducting correlations with single-band
ladders we have evaluated the pair-pair correlation function $P(r) =
\langle P_i^\dagger P_j \rangle$, with $r=|i-j|$, where $P_i^\dagger$
is the Cu-Cu ladder rung spin-singlet (see Fig.~\ref{fig-pairs}(a)),
defined as
$2^{-1/2}(d^\dagger_{i,1,\uparrow}d^\dagger_{i,2,\downarrow}-d^\dagger_{i,1,\downarrow}d^\dagger_{i,2,\uparrow})$.
Figs.~\ref{fig-length}(a) and (b) show the ladder length dependence of
$P(r)$ for two different parameters sets, Fig.~\ref{fig-length}(a)
with $U_{\rm p} =t_{\rm pp}=0$ and Fig.~\ref{fig-length}(b) with
$U_{\rm p}=3$ and $t_{\rm pp}=0.5$. For all of the ladder lengths and
parameter values we studied, $P(r)$ was well fit by a power law
$r^{-\alpha}$, provided short and long distances are excluded. These
limits are due to finite size effects and are well understood in the
case of the one-band ladder \cite{Dolfi15a}. In particular, the sharp
decrease in $P(r)$ for $r > L/2$ in Figs.~\ref{fig-length} and
\ref{fig-96rung} is a finite-size effect and not due to insufficiently
large DMRG $m$.  As seen in Fig.~\ref{fig-length}(a), for $U_{\rm p}
=t_{\rm pp}=0$ we find $\alpha\sim 1$.  With more realistic
parameters, $U_{\rm p}=3$ and $t_{\rm pp}=0.5$, we find the power law
exponent $\alpha$ close to 1.5 (see Fig.~\ref{fig-length}(b)).

In Fig.~\ref{fig-96rung} we show power-law fits for $\delta=0.0625$
(Fig.~\ref{fig-96rung}(a)) and $\delta=0.125$
(Fig.~\ref{fig-96rung}(b)) for the 96-rung ladder.  Nonzero $U_{\rm
  p}$ and $t_{\rm pp}$ both suppress $P(r)$, and when {\it both} are
nonzero the suppression of $P(r)$ is further increased.  This is
consistent with the synergistic suppression of the spin gap in the
doped ladder.  As in the one-band ladder, we find that $\alpha$
increases rapidly with doping \cite{Dolfi15a}.

\begin{figure}
  \centerline{\resizebox{3.4in}{!}{\includegraphics{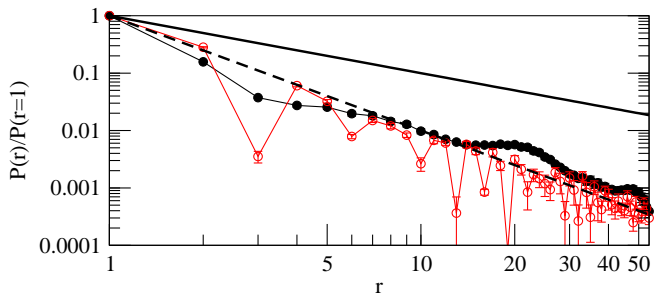}}}
  \caption{(Color online) Rung-singlet $P(r)$ for the 64-rung lattice
    with $U_{\rm d}=8$, $U_{\rm p}=3$, $t_{\rm pp}$=0.5, and
    $\delta=0.0625$. $P(r)$ is normalized by its value at $r=1$. Solid
    (open) symbols are for lattice of Fig.~\ref{fig-lattice}(a)
    (Fig.~\ref{fig-lattice}(b)).  The solid (dashed) lines are power
    laws $r^{-1}$ ($r^{-2}$).}
  \label{fig-bothlattice}
\end{figure}

As mentioned above, we have also performed calculations of the rung
singlet superconducting pair correlations for the ladder with the
geometry of Fig.~\ref{fig-lattice}(b). The results of these
calculations are shown in Fig.~\ref{fig-bothlattice}, where we compare
the distance dependences of the normalized rung singlet pair-pair
correlations for the two geometries of Fig.~\ref{fig-lattice}.  The
distance dependence of the pair correlations for the geometry of
Fig.~\ref{fig-lattice}(b) with outer O ions included is as rapid as
that for geometry of Fig.~\ref{fig-lattice}(a). As we point out below,
this result is to be anticipated from physical reasonings.

\subsubsection{Other pairing symmetries}

\begin{figure}
  \centerline{\resizebox{3.4in}{!}{\includegraphics{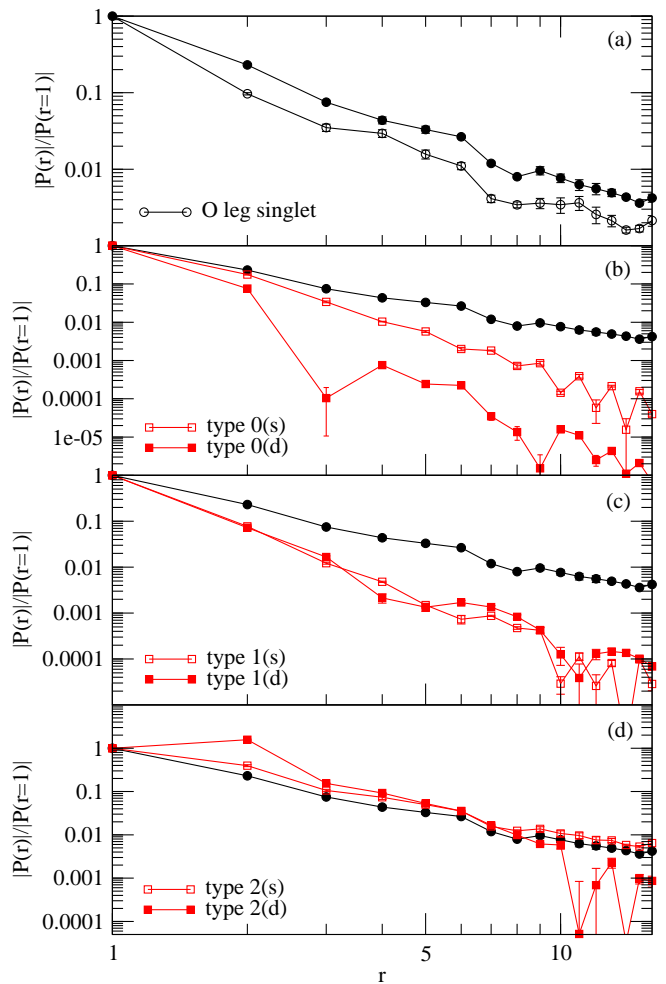}}}
  \caption{(Color online) Pair-pair correlations for
      32-rung ladders of Fig.~\ref{fig-lattice}(a) for the
      oxygen based pair symmetries in Fig.~\ref{fig-pairs}. Parameters are
      $U_{\rm d}=8$, $U_{\rm p}=3$, $t_{\rm pp}=0.5$, and
      $\delta=0.125$.  All $P(r)$ are normalized by their $r=1$ value;
      we have taken absolute values of correlations that are negative.
      (a) pair symmetry of Fig.~\ref{fig-pairs}(b); (a) pair symmetry
      of Fig.~\ref{fig-pairs}(c); (c) pair symmetry of
      Fig.~\ref{fig-pairs}(d); (d) pair symmetry of
      Fig.~\ref{fig-pairs}(e); Round symbols are for Cu rung pairs
      (Fig.~\ref{fig-pairs}(a)), square open (filled) symbols are for
      $s$ and $d$-wave pairing of the specified type (see text).}
  \label{fig-othercorrel1}
\end{figure}

\begin{figure}
  \centerline{\resizebox{3.4in}{!}{\includegraphics{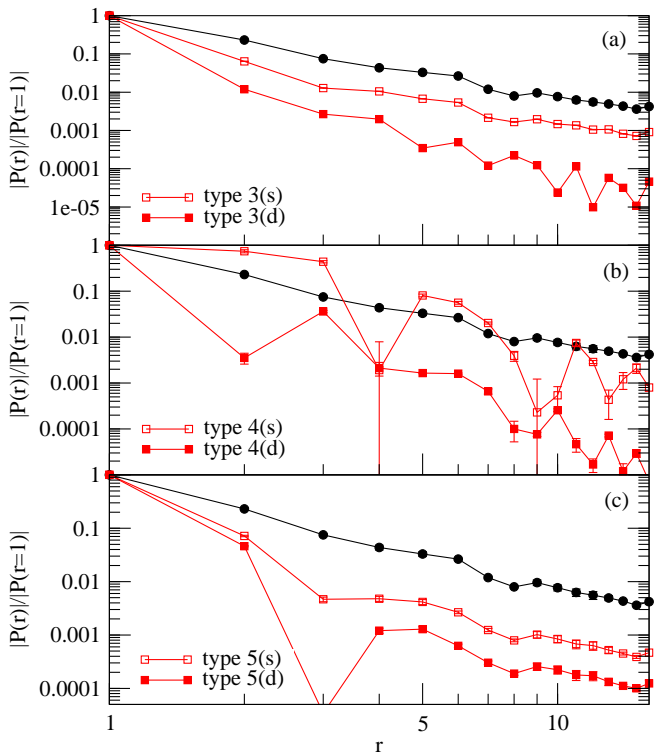}}}
  \caption{(Color online) Pair-pair correlations for
      32-rung ladders of Fig.~\ref{fig-lattice}(a) for the
      copper based pair symmetries in Fig.~\ref{fig-pairs}. Hamiltonian
      parameters are identical to Fig.~\ref{fig-othercorrel1}. All
      $P(r)$ are normalized by their $r=1$ value; we have taken
      absolute values of correlations that are negative.  (a) pair
      symmetry of Fig.~\ref{fig-pairs}(f); (a) pair symmetry of
      Fig.~\ref{fig-pairs}(g); (c) pair symmetry of
      Fig.~\ref{fig-pairs}(h).  Round symbols are for Cu rung pairs
      (Fig.~\ref{fig-pairs}(a)), square open (filled) symbols are for
      $s$ and $d$-wave pairing of the specified type (see text).}
  \label{fig-othercorrel2}
\end{figure}

Given that the doped holes primarily occupy the O$_R$ and O$_L$ sites
investigation of pair-pair correlations beyond those involving simple
rung singlets is important (note that the overall charges continue to
reside primarily on the Cu-ions though).  We investigated several
different pair symmetries composed of superpositions of singlets on Cu
and O atoms. The pair symmetries we investigated are shown in
Fig.~\ref{fig-pairs}. We have calculated $P(r \equiv |i-j|) = \langle
P_i^\dagger P_j \rangle$ for these different pairing symmetries for a
32-rung ladder with $U_{\rm d} = 8$, $\Delta_{\rm dp}=3$, $U_{\rm
  p}=3$ and $t_{\rm pp}=0.5$, for doping $\delta=0.125$. Our results
are summarized in Figs.~\ref{fig-othercorrel1} and
\ref{fig-othercorrel2}, where we compare the distance dependences of
these pair correlations with the original pair correlation involving
rung singlets only (Fig.~\ref{fig-pairs}(a).  As seen in Figs
\ref{fig-othercorrel1} and \ref{fig-othercorrel2}, all of the pairing
symmetries we investigated fall into one of two categories:

1. $P(r)$ has identical long distance decay with $P(r)$ for the rung
singlet correlation (Fig.~\ref{fig-pairs}(a)). This includes
Fig.~\ref{fig-pairs}(b), Fig.~\ref{fig-pairs}(e) (s-wave only),
Fig.~\ref{fig-pairs}(f) (s-wave only) and
Fig.~\ref{fig-pairs}(h). These pair symmetries either contain within
their superposition of pairs the rung singlet itself, or in case of
Fig.~\ref{fig-pairs}(b) a singlet between O atoms immediately adjacent
to the Cu rung atoms.

2. $P(r)$ decays much faster with distance than the rung singlet
correlation. This includes Figs.~\ref{fig-pairs}(c) and (d),
Fig.~\ref{fig-pairs}(e) (d-wave only), Fig.~\ref{fig-pairs}(f) (d-wave
only) and Fig.~\ref{fig-pairs}(g).

Based on these results we conclude that superconducting pair-pair
correlations in the three-band ladder in general decay much faster
than in the single-band ladder, and for realistic correlation and
hopping parameters this decay precludes quasi-long range order.

\subsection{Density oscillations}
\label{dens}

In the Luther-Emery phase, the long-distance decay of pair-pair
correlations follows a power law determined by the correlation
parameter $\kappa_\rho$, with $P(r) \propto r^{-1/\kappa_\rho}$
($\alpha=1/\kappa_\rho$).  Similarly, density-density correlations
$N(r)$ follow a power law with $N(r) \propto r^{-\kappa_\rho}$
($N(r=|i-j|)=\langle n_i n_j \rangle - \langle n_i\rangle\langle
n_j\rangle$, where $n_i$ is the charge density operator for site $i$).
Pairing correlations decay slower than $1/r$ and dominate over
density-density correlations only for $\kappa_\rho>1$. Density
correlations provide a second estimate of $\kappa_\rho$ and
consistency check.
  
In the 2-leg one-band ladder, charge density (Friedel) oscillations
can be fit to accurately extract $\kappa_\rho$
\cite{White02a,Dolfi15a}. We use a similar procedure in our three-band
model calculations. We fit the charge density to the following
function \cite{White02a,Dolfi15a}
  \begin{figure}[t]
  \centerline{\resizebox{3.4in}{!}{\includegraphics{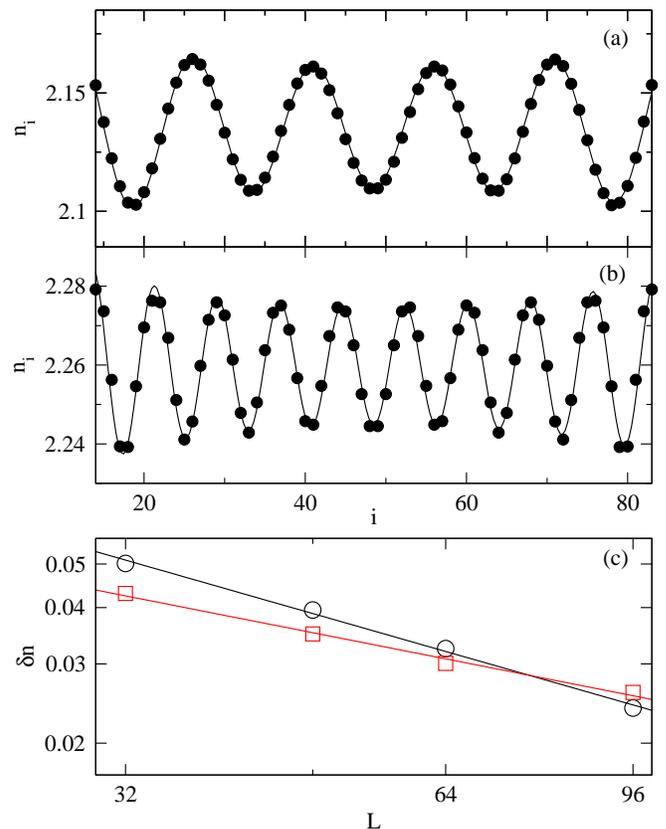}}}
  \caption{(Color online) (a) Total charge density versus rung
    index $i$ for a 96-rung ladder with $U_{\rm p}=4$, $t_{\rm pp}=0.6$,
    and $\delta=0.0625$. The line is a fit to Eq.~\ref{eqn-dens}.
    (b) Same as (a) but $\delta=0.125$. (c) Amplitude of density
    oscillations at $L/2$ versus ladder length. Circles (squares)
    are for $U_{\rm p}=0$ ($U_{\rm p}=4$), $t_{\rm pp}=0$ ($t_{\rm pp}=0.6$),
    and $\delta=0.0625$. Lines are fits (see text).}
  \label{fig-density}
\end{figure}
\begin{equation}
  n_j = A \frac{\cos(\pi N_h j/L_{\rm eff}+\phi)}{\sin(\pi j/L)^{\kappa_\rho/2}}+n_0.
  \label{eqn-dens}
\end{equation}  
In Eq.~\ref{eqn-dens}, $n_j$ is the charge density in the $j$th
unit cell
of the lattice; $n_j$ is the sum of the charge densities of
the rung copper and oxygen atoms, plus the charge densities on two
adjacent leg oxygen atoms. $N_h$ is the number of holes, $A$ the
overall amplitude, $\phi$ a phase shift, and $n_0$ the background
charge density. $L_{\rm eff}$ is an effective length that is less than
$L$ because of the finite extent of holes \cite{Dolfi15a}. We
performed a nonlinear fit of the charge density to Eq.~\ref{eqn-dens}
in the central region of the ladders for $L$=32, 48, 64, and 96,
keeping $A$, $n_0$, $\kappa_\rho$, and $\phi$ as free parameters.
Fig.~\ref{fig-density}(a) and (b) show representative fits for
$L=96$. While in the one-band ladder a good fit was found
\cite{Dolfi15a} with $L_{\rm eff}= L-2$, we found a shorter $L_{\rm
  eff}$ in the three-band ladder, for example $L_{\rm eff}\approx L-9$
and $L-6$ for $\delta=0.0625$ and $\delta=0.125$, respectively. This
shows that doped holes extend over a larger number of lattice sites
in the three-band ladder than in the one-band ladder (see
also Section \ref{physical-picture}).

The value of $\kappa_\rho$ can be most accurately determined in
ladders \cite{Dolfi15a} from the scaling of the amplitude of the density
oscillations in the center of the ladder,
\begin{equation}
  \delta n(L)=n(L/2)-n_0 \propto L^{-\kappa_\rho/2}.
  \label{deltan}
\end{equation}
In Eq.~\ref{deltan} we determined $n(L/2)$ and $n_0$ from the fit of
Eq.~\ref{eqn-dens}.  Fig.~\ref{fig-density}(c) shows typical results
for $\delta n$ for two different parameter sets (see
\cite{Supplemental} for other parameter values).  We then estimated
$\kappa_\rho$ in the $L\rightarrow\infty$ limit from a linear fit as
was done for the one-band ladder \cite{Dolfi15a}. Table
\ref{krho-table} summarizes our results for $\kappa_\rho$ as
determined from directly fitting $P(r)$ and from Eq.~\ref{deltan}. In
comparison, the value of $\kappa_\rho$ in the single band ladder is
1.54--1.66 at $\delta=0.0625$ and 0.92-1.17 at $\delta=0.125$
\cite{Dolfi15a}, roughly consistent with our $U_{\rm p}=t_{\rm pp}=0$
values.  In general, the $L\rightarrow\infty$ $\kappa_\rho$ determined
from density oscillations is larger than the value found from the
$L=96$ pair-pair correlations, with a larger finite-size correction at
$\delta=0.0625$ than at $\delta=0.125$.  A slight decrease in the
power-law slope of $P(r)$ with increasing size is evident in
Fig.~\ref{fig-length}, so we believe that the results from fitting
$P(r)$ and $\delta n$ are consistent with each other.

We note that the $\kappa_\rho$ calculated from the two different
methods display {\it nearly identical} dependence on $U_{\rm p}$ and
$t_{\rm pp}$.  The effects of nonzero $U_{\rm p}$ and $t_{\rm pp}$ are
not simply additive. Rather (see Table \ref{krho-table}), the decrease
of $\kappa_\rho$ is larger when {\it both} $U_{\rm p}$ and $t_{\rm
  pp}$ are nonzero compared to the sum of the changes in $\kappa_\rho$
with ($U_{\rm p}>0$, $t_{\rm pp}=0$) or ($U_{\rm p}=0$, $t_{\rm
  pp}>0$).  We note that we cannot rule out the possibility that
$\kappa_\rho>1$ in the limit of very small doping; indeed we find
$\kappa_\rho\sim 1$ for $\delta=0.0625$ even with realistic $U_{\rm
  p}$ and $t_{\rm pp}$.  However, if one takes the doping typically
assumed as maximizing pairing in the 2D one-band Hubbard model
($\delta=0.125$), we find that $\kappa_\rho$ is significantly less
than one, even considering our computational and finite-size errors.
This indicates the absence of the quasi-long range SC characteristic
of the single-band ladder.

\begin{table}
\begin{ruledtabular}
\begin{tabular}{ccccc}
  $\delta$ & $U_{\rm p}$ & $t_{\rm pp}$ & \multicolumn{2}{c}{$\kappa_\rho$}  \\
           &            &            & $P(r)$ ($L$=96) & $\delta n$ ($L\rightarrow\infty$) \\
  \hline
  0.0625 & 0.0 & 0.0 & 1.14(4) & 1.35(6)  \\
  & 0.0 & 0.5 & 0.99(3) &  1.22(2) \\
  & 3.0 & 0.0 & 1.01(3) & 1.20(6) \\
  & 3.0 & 0.5 & 0.76(2) &  0.97(1) \\
  & 4.0 & 0.6 & 0.65(2) & 0.93(5) \\ 
  0.125 & 0.0 & 0.0 & 0.91(2) &  1.08(6) \\
  & 0.0 & 0.5 & 0.78(2) & 0.93(2)  \\
  & 3.0 & 0.0 & 0.83(2) & 1.08(3)  \\
  & 3.0 & 0.5 & 0.65(2) & 0.80(4)  \\
  & 4.0 & 0.6 & 0.56(2) & 0.75(8)  \\
  \hline
\end{tabular}
\end{ruledtabular}
\caption{Correlation exponents $\kappa_\rho$ obtained
  from fitting $P(r)$ (Section \ref{pr}) and density oscillations
  (Section \ref{dens}).}
\label{krho-table}
\end{table}

\subsection{Suppression of pair correlations, a physical picture}
\label{physical-picture}

The suppression of the pair-pair correlations within the multi-band
model is reminiscent of similar suppression of the same correlations
for large rung hopping $t_{\perp}>t$ within the one-band model (see
Fig.~6 in reference \onlinecite{Noack97a}).  As seen in reference
\onlinecite{Noack97a} not only is the one-band pair-pair correlation
suppressed by large t$_{\perp}$, the suppression occurs at smaller and
smaller $t_{\perp}$ (that are however $>1$) as the Hubbard repulsion
$U$ increases.  Within the same range of $U$ the spin gap in the
undoped one-band ladder {\it increases} with $U$.  It therefore
follows that {\it increase in the spin gap in the undoped one-band
  ladder is accompanied by concomitant increase of pair correlations
  in the doped ladder, only until a maximum in the undoped ladder spin
  gap is reached.} Beyond this maximum, further increase of the spin
gap in the undoped single-band ladder results in suppression of the
pair-pair correlations in the doped ladder. {\it Our results in
  Figs.~\ref{fig-96rung}-\ref{fig-othercorrel2} indicate that this
  maximum in the spin gap of the undoped three-band ladder has been
  reached already at $t_{\rm dp}^\perp=1$.} We argue in the following
that this is due to the large pair-breaking effect in the multi-band
ladder.

Superconducting pairing involving rung singlets in both one- and
multi-band ladder Hamiltonians can understood within an {\it
  effective} Hamiltonian of the form,
\begin{equation}
H_{\rm eff}= \sum_i J(\delta)P_i^\dagger P_i-t_{\rm pair}\sum_{\langle i,j \rangle }P_i^\dagger P_j-
t_{\rm f}\sum_{\langle \mu,\nu \rangle, \sigma} f_{\mu,\sigma}^\dagger f_{\nu,\sigma}
\end{equation}  
where $J(\delta)$ is proportional to the self-consistent spin gap at
doping $\delta$, $t_{\rm pair}$ the effective pair hopping integral,
and $t_{\rm f}$ refers to single-particle fermion hops.  Here $i$ and
$j$ refer to rung indices. While $\mu$,$\nu$ refer to Cu-ions on
nearest neighbor rungs in the one-band ladder, they refer to both Cu-
and O-sites in the multi-band ladder.  Within the one-band model,
$t_{\rm pair}$ and $t_{\rm f}$ are related, with $t_{\rm pair} \sim
t_{\rm f}^2/\Delta_{\rm pb}$, where $\Delta_{\rm pb}$ is the
pair-binding energy, roughly proportional to the spin gap in the doped
ladder.

The interactions $J(\delta)$ and $t_{\rm pair}$, taken together,
dominate over the pair-breaking single-particle $t_{\rm f}$ over a
broad range of parameters in the one-band ladder, including in
particular $t_{\perp}=1$. This situation is altered significantly
within the multi-band model. The doped holes now enter primarily
O-sites (see Table 1), and the complete spin-singlet wavefunction
involves not only the Cu-ions but also the rung O-ion and the four
ladder oxygens on either side of the rung.  Pair motion now must
involve not only the doped charges on the Cu-ions of a rung, but also
those on the neighboring O-ions, making the effective mass of the
spin-singlet within the three-band model considerably larger than in
the one-band model.  At the same time, however, $t_{\rm f}$ now can
involve the holes on the O-ions exclusively, with the Cu-ion holes
playing a very limited role ({\it i.e.}, $t_{\rm f}$ now includes and
is even dominated by $t_{\rm pp}$). Consequently, the effective
$t_{\rm pair}$ is smaller and the effective $t_{\rm f}$ larger within
the three-band model.  {\it Single-particle hopping thus has a far
  stronger pair-breaking effect in the three-band ladder.}

 Based on the above it now becomes obvious why the strongest
 suppression of the doped-state spin gap and rung singlet
 superconducting pair correlations occur within the Hamiltonians with
 nonzero $t_{\rm pp}$ (see Figs.~\ref{fig-gaps}(b) and
 \ref{fig-96rung}).  With the geometry of Fig.~\ref{fig-lattice}(b),
 the effective mass of the spin-singlet is further enhanced while the
 additional $t_{\rm pp}$ contribute to additional $t_{\rm f}$ and
 pair-breaking. The strong suppression of pair correlations should
 therefore be common to the geometries of Fig.~\ref{fig-lattice}(a)
 and (b), as indeed is found numerically.

\section{Discussions and Conclusion}

Our theoretical results demonstrate that, (i) conclusions regarding
pairing based on effective single-band ladder models cannot be
extended to the three-band ladder, and (ii) there is no pairing within
the three-band ladder for realistic cuprate parameters $U_{\rm
  p}=3-4$, $t_{\rm pp}=0.5-0.6$ \cite{Hirayama18a}.  $U_{\rm p}$ and
$t_{\rm pp}$ both suppresses pairing uniformly.

The absence of SC in the 2-leg ladder compound
La$_{2-x}$Sr$_x$CuO$_{2.5}$ \cite{Hiroi95a} is therefore expected
within our theory.  Superconducting Sr$_{14-x}$Ca$_x$Cu$_{24}$O$_{41}$
consists of alternating planes of corner-sharing CuO$_2$ chains and
edge-sharing Cu$_2$O$_3$ ladders
\cite{Uehara96a,Nagata98a,Vuletic06a}.  It is believed that there
occur nearly 5 holes per formula unit (f. u.) on chains and 1 hole per
f. u. on ladders at $x=0$.  There occurs some transfer of holes from
chain to ladder with increasing $x$, but the actual extent of the
transfer is not agreed upon \cite{Bugnet16a}.  The appearance of SC
above 4.0 GPa in $x=11.5$ single crystals is accompanied by a 1D-to-2D
dimensional crossover, as evidenced from the insulator-like
resistivity $\rho_a$ along the rung-axis $a$ at all temperatures below
the critical pressure and metallic $\rho_a$ at all temperatures above
this pressure \cite{Nagata98a}.  The resistivity ratio $\rho_a/\rho_c$
(the $c$-axis corresponds to the ladder leg direction) of the $x=11.5$
compound decreases by more than a factor of 4 at low temperature and
high pressure \cite{Nagata98a}. There occurs a concomitant decrease in
the $a$-axis lattice parameter, although at still higher pressure
where superconducting T$_c$ decreases the lattice parameter increases
again.  $^{63}$Cu and $^{17}$O NMR studies for the $x=12$ compound
have found that the spin gap decreases sharply with pressure, and
there appear low-lying spin excitations, indicating the presence of
mobile quasi-particles that contribute to a finite density of states
at the Fermi level and perhaps also SC \cite{Piskunov05a,Fujiwara09a}.
Taken together, these observations, (i) indicate clearly that the
origin of SC in Sr$_{14-x}$Ca$_x$Cu$_{24}$O$_{41}$ cannot be found
within ladder-based theories \cite{Vuletic06a}, and (ii) are
consistent with our finding that superconducting correlations are
absent in the three-band ladder Hamiltonian with realistic $U_{\rm p}$
and $t_{\rm pp}$.

Our results raise a fundamental (and disturbing) question.  What is
the implication of the absence of a Luther-Emery superconducting phase
\cite{Luther74a} in the three-band 2-leg ladder Hubbard Hamiltonian
for the 2D CuO$_2$ layer? We make the following observations.  First,
theories of cuprate SC that assume a gapped spin-liquid phase
proximate to the superconducting state
\cite{Anderson87b,Emery97a,Lee06b,Arrigoni04a,Hur09a} cannot be
justified by the demonstration of quasi-long range superconducting
correlations within the one-band ladder-based theories. The one-band
ladder model is an artificial one with no relationship to real
cuprates.  Second, the profound difference between the results of one-
and three-band ladder calculations (understandable physically with
hindsight, see Section III.D) suggests that the mapping of the
three-band Hamiltonian to the one-band Hubbard model \cite{Zhang88a}
is correct only for a limited choice of parameters. In the context of
cuprates, the applicability of the mapping across doping levels,
realistic geometries and parameters (especially $t_{\rm pp}$) has been
questioned by other authors \cite{Peets09a,Sunko09a,Adolphs16a},
although these criticisms themselves remain controversial.  Our work
suggests that re-evaluation of these earlier works is
necessary. Finally, as with the 2D one-band Hubbard Hamiltonian,
numerical computation of $d_{x^2-y^2}$ pair correlations within the
three-band Hamiltonian for the CuO$_2$ layer also found absence of SC
\cite{Guerrero98a}. This latter work used the constrained path quantum
Monte Carlo approach that relies on a trial wavefunction to eliminate
the Fermion sign problem. The calculations were also based on
relatively small lattices. Our DMRG calculations, devoid of sign
errors, provide strong support to the conclusions of reference
\onlinecite{Guerrero98a}. Taken together, these observations suggest
that a comprehensive theory of cuprate SC may require starting
hypotheses or models that are significantly different from existing
ones.

Work at Arizona was supported by NSF-CHE-1764152.  Some calculations
in this work were supported under project TG-DMR190068 of the Extreme
Science and Engineering Discovery Environment \cite{xsede} (XSEDE),
which is supported by National Science Foundation grant number
ACI-1548562.  Specifically, we used the Bridges and Bridges2 systems
at the Pittsburgh Supercomputing Center, which are supported by NSF
awards ACI-1445606 and ACI-1928147, respectively.


\begin{thebibliography}{44}%
\makeatletter
\providecommand \@ifxundefined [1]{%
 \@ifx{#1\undefined}
}%
\providecommand \@ifnum [1]{%
 \ifnum #1\expandafter \@firstoftwo
 \else \expandafter \@secondoftwo
 \fi
}%
\providecommand \@ifx [1]{%
 \ifx #1\expandafter \@firstoftwo
 \else \expandafter \@secondoftwo
 \fi
}%
\providecommand \natexlab [1]{#1}%
\providecommand \enquote  [1]{``#1''}%
\providecommand \bibnamefont  [1]{#1}%
\providecommand \bibfnamefont [1]{#1}%
\providecommand \citenamefont [1]{#1}%
\providecommand \href@noop [0]{\@secondoftwo}%
\providecommand \href [0]{\begingroup \@sanitize@url \@href}%
\providecommand \@href[1]{\@@startlink{#1}\@@href}%
\providecommand \@@href[1]{\endgroup#1\@@endlink}%
\providecommand \@sanitize@url [0]{\catcode `\\12\catcode `\$12\catcode
  `\&12\catcode `\#12\catcode `\^12\catcode `\_12\catcode `\%12\relax}%
\providecommand \@@startlink[1]{}%
\providecommand \@@endlink[0]{}%
\providecommand \url  [0]{\begingroup\@sanitize@url \@url }%
\providecommand \@url [1]{\endgroup\@href {#1}{\urlprefix }}%
\providecommand \urlprefix  [0]{URL }%
\providecommand \Eprint [0]{\href }%
\providecommand \doibase [0]{https://doi.org/}%
\providecommand \selectlanguage [0]{\@gobble}%
\providecommand \bibinfo  [0]{\@secondoftwo}%
\providecommand \bibfield  [0]{\@secondoftwo}%
\providecommand \translation [1]{[#1]}%
\providecommand \BibitemOpen [0]{}%
\providecommand \bibitemStop [0]{}%
\providecommand \bibitemNoStop [0]{.\EOS\space}%
\providecommand \EOS [0]{\spacefactor3000\relax}%
\providecommand \BibitemShut  [1]{\csname bibitem#1\endcsname}%
\let\auto@bib@innerbib\@empty
%</preamble>
\bibitem [{\citenamefont {Anderson}\ \emph {et~al.}(1987)\citenamefont
  {Anderson}, \citenamefont {Baskaran}, \citenamefont {Zou},\ and\
  \citenamefont {Hsu}}]{Anderson87b}%
  \BibitemOpen
  \bibfield  {author} {\bibinfo {author} {\bibfnamefont {P.~W.}\ \bibnamefont
  {Anderson}}, \bibinfo {author} {\bibfnamefont {G.}~\bibnamefont {Baskaran}},
  \bibinfo {author} {\bibfnamefont {Z.}~\bibnamefont {Zou}},\ and\ \bibinfo
  {author} {\bibfnamefont {T.}~\bibnamefont {Hsu}},\ }\bibfield  {title}
  {\bibinfo {title} {Resonating valence-bond theory of phase transitions and
  superconductivity in \protect{La$_2$CuO$_4$}-based compounds},\ }\href@noop
  {} {\bibfield  {journal} {\bibinfo  {journal} {Phys.\ Rev.\ Lett.}\ }\textbf
  {\bibinfo {volume} {58}},\ \bibinfo {pages} {2790} (\bibinfo {year}
  {1987})}\BibitemShut {NoStop}%
\bibitem [{\citenamefont {Anderson}\ \emph {et~al.}(2004)\citenamefont
  {Anderson}, \citenamefont {Lee}, \citenamefont {Randeria}, \citenamefont
  {Rice}, \citenamefont {Trivedi},\ and\ \citenamefont {Zhang}}]{Anderson04a}%
  \BibitemOpen
  \bibfield  {author} {\bibinfo {author} {\bibfnamefont {P.~W.}\ \bibnamefont
  {Anderson}}, \bibinfo {author} {\bibfnamefont {P.~A.}\ \bibnamefont {Lee}},
  \bibinfo {author} {\bibfnamefont {M.}~\bibnamefont {Randeria}}, \bibinfo
  {author} {\bibfnamefont {T.~M.}\ \bibnamefont {Rice}}, \bibinfo {author}
  {\bibfnamefont {N.}~\bibnamefont {Trivedi}},\ and\ \bibinfo {author}
  {\bibfnamefont {F.~C.}\ \bibnamefont {Zhang}},\ }\bibfield  {title} {\bibinfo
  {title} {The physics behind high-temperature superconducting cuprates: the
  `plain vanilla' version of {R}{V}{B}},\ }\href@noop {} {\bibfield  {journal}
  {\bibinfo  {journal} {J. Phys. Condens. Matter}\ }\textbf {\bibinfo {volume}
  {16}},\ \bibinfo {pages} {R755} (\bibinfo {year} {2004})}\BibitemShut
  {NoStop}%
\bibitem [{\citenamefont {Lee}\ \emph {et~al.}(2006)\citenamefont {Lee},
  \citenamefont {Nagaosa},\ and\ \citenamefont {Wen}}]{Lee06b}%
  \BibitemOpen
  \bibfield  {author} {\bibinfo {author} {\bibfnamefont {P.~A.}\ \bibnamefont
  {Lee}}, \bibinfo {author} {\bibfnamefont {N.}~\bibnamefont {Nagaosa}},\ and\
  \bibinfo {author} {\bibfnamefont {X.~G.}\ \bibnamefont {Wen}},\ }\bibfield
  {title} {\bibinfo {title} {Doping a {M}ott insulator: {P}hysics of
  high-temperature superconductivity},\ }\href@noop {} {\bibfield  {journal}
  {\bibinfo  {journal} {Rev.\ Mod.\ Phys.}\ }\textbf {\bibinfo {volume} {78}},\
  \bibinfo {pages} {17} (\bibinfo {year} {2006})}\BibitemShut {NoStop}%
\bibitem [{\citenamefont {Scalapino}(2012)}]{Scalapino12a}%
  \BibitemOpen
  \bibfield  {author} {\bibinfo {author} {\bibfnamefont {D.~J.}\ \bibnamefont
  {Scalapino}},\ }\bibfield  {title} {\bibinfo {title} {A common thread: The
  pairing interaction for unconventional superconductors},\ }\href@noop {}
  {\bibfield  {journal} {\bibinfo  {journal} {Rev.\ Mod.\ Phys.}\ }\textbf
  {\bibinfo {volume} {84}},\ \bibinfo {pages} {1383} (\bibinfo {year}
  {2012})}\BibitemShut {NoStop}%
\bibitem [{\citenamefont {Zhang}\ and\ \citenamefont {Rice}(1988)}]{Zhang88a}%
  \BibitemOpen
  \bibfield  {author} {\bibinfo {author} {\bibfnamefont {F.~C.}\ \bibnamefont
  {Zhang}}\ and\ \bibinfo {author} {\bibfnamefont {T.~M.}\ \bibnamefont
  {Rice}},\ }\bibfield  {title} {\bibinfo {title} {Effective {H}amiltonian for
  the superconducting {C}u oxides},\ }\href@noop {} {\bibfield  {journal}
  {\bibinfo  {journal} {Phys.\ Rev.\ B}\ }\textbf {\bibinfo {volume} {37}},\
  \bibinfo {pages} {3759(R)} (\bibinfo {year} {1988})}\BibitemShut {NoStop}%
\bibitem [{\citenamefont {Zhang}\ \emph {et~al.}(1997)\citenamefont {Zhang},
  \citenamefont {Carlson},\ and\ \citenamefont {Gubernatis}}]{Zhang97b}%
  \BibitemOpen
  \bibfield  {author} {\bibinfo {author} {\bibfnamefont {S.}~\bibnamefont
  {Zhang}}, \bibinfo {author} {\bibfnamefont {J.}~\bibnamefont {Carlson}},\
  and\ \bibinfo {author} {\bibfnamefont {J.~E.}\ \bibnamefont {Gubernatis}},\
  }\bibfield  {title} {\bibinfo {title} {Pairing correlations in the
  two-dimensional {H}ubbard model},\ }\href@noop {} {\bibfield  {journal}
  {\bibinfo  {journal} {Phys.\ Rev.\ Lett.}\ }\textbf {\bibinfo {volume}
  {78}},\ \bibinfo {pages} {4486} (\bibinfo {year} {1997})}\BibitemShut
  {NoStop}%
\bibitem [{\citenamefont {Aimi}\ and\ \citenamefont {Imada}(2007)}]{Aimi07a}%
  \BibitemOpen
  \bibfield  {author} {\bibinfo {author} {\bibfnamefont {T.}~\bibnamefont
  {Aimi}}\ and\ \bibinfo {author} {\bibfnamefont {M.}~\bibnamefont {Imada}},\
  }\bibfield  {title} {\bibinfo {title} {Does simple two-dimensional {H}ubbard
  model account for high-\protect{T$_c$} superconductivity in copper oxides?},\
  }\href@noop {} {\bibfield  {journal} {\bibinfo  {journal} {J.\ Phys.\ Soc.\
  Jpn.}\ }\textbf {\bibinfo {volume} {76}},\ \bibinfo {pages} {113708}
  (\bibinfo {year} {2007})}\BibitemShut {NoStop}%
\bibitem [{\citenamefont {Qin}\ \emph {et~al.}(2020)\citenamefont {Qin},
  \citenamefont {Chung}, \citenamefont {Shi}, \citenamefont {Vitali},
  \citenamefont {Hubig}, \citenamefont {\protect{Schollw\"ock}}, \citenamefont
  {White},\ and\ \citenamefont {Zhang}}]{Qin20a}%
  \BibitemOpen
  \bibfield  {author} {\bibinfo {author} {\bibfnamefont {M.}~\bibnamefont
  {Qin}}, \bibinfo {author} {\bibfnamefont {C.-M.}\ \bibnamefont {Chung}},
  \bibinfo {author} {\bibfnamefont {H.}~\bibnamefont {Shi}}, \bibinfo {author}
  {\bibfnamefont {E.}~\bibnamefont {Vitali}}, \bibinfo {author} {\bibfnamefont
  {C.}~\bibnamefont {Hubig}}, \bibinfo {author} {\bibfnamefont
  {U.}~\bibnamefont {\protect{Schollw\"ock}}}, \bibinfo {author} {\bibfnamefont
  {S.~R.}\ \bibnamefont {White}},\ and\ \bibinfo {author} {\bibfnamefont
  {S.}~\bibnamefont {Zhang}},\ }\bibfield  {title} {\bibinfo {title} {Absence
  of superconductivity in the pure two-dimensional {H}ubbard model},\
  }\href@noop {} {\bibfield  {journal} {\bibinfo  {journal} {Phys. Rev. X}\
  }\textbf {\bibinfo {volume} {10}},\ \bibinfo {pages} {031016} (\bibinfo
  {year} {2020})}\BibitemShut {NoStop}%
\bibitem [{\citenamefont {Huang}\ \emph {et~al.}(2001)\citenamefont {Huang},
  \citenamefont {Lin},\ and\ \citenamefont {Gubernatis}}]{Huang01a}%
  \BibitemOpen
  \bibfield  {author} {\bibinfo {author} {\bibfnamefont {Z.~B.}\ \bibnamefont
  {Huang}}, \bibinfo {author} {\bibfnamefont {H.~Q.}\ \bibnamefont {Lin}},\
  and\ \bibinfo {author} {\bibfnamefont {J.~E.}\ \bibnamefont {Gubernatis}},\
  }\bibfield  {title} {\bibinfo {title} {Quantum {M}onte {C}arlo study of spin,
  charge, and pairing correlations in the {t}-{t$^\prime$}-{$U$} {H}ubbard
  model},\ }\href@noop {} {\bibfield  {journal} {\bibinfo  {journal} {Phys.\
  Rev.\ B}\ }\textbf {\bibinfo {volume} {64}},\ \bibinfo {pages} {205101}
  (\bibinfo {year} {2001})}\BibitemShut {NoStop}%
\bibitem [{\citenamefont {Chung}\ \emph {et~al.}(2020)\citenamefont {Chung},
  \citenamefont {Qin}, \citenamefont {Zhang}, \citenamefont
  {\protect{Schollw\"ock}},\ and\ \citenamefont {White}}]{Chung20a}%
  \BibitemOpen
  \bibfield  {author} {\bibinfo {author} {\bibfnamefont {C.-M.}\ \bibnamefont
  {Chung}}, \bibinfo {author} {\bibfnamefont {M.}~\bibnamefont {Qin}}, \bibinfo
  {author} {\bibfnamefont {S.}~\bibnamefont {Zhang}}, \bibinfo {author}
  {\bibfnamefont {U.}~\bibnamefont {\protect{Schollw\"ock}}},\ and\ \bibinfo
  {author} {\bibfnamefont {S.~R.}\ \bibnamefont {White}},\ }\bibfield  {title}
  {\bibinfo {title} {Plaquette versus ordinary $d$-wave pairing in the
  $t^\prime$-{H}ubbard model on a width 4 cylinder},\ }\href@noop {} {\bibfield
   {journal} {\bibinfo  {journal} {Phys.\ Rev.\ B}\ }\textbf {\bibinfo {volume}
  {102}},\ \bibinfo {pages} {041106(R)} (\bibinfo {year} {2020})}\BibitemShut
  {NoStop}%
\bibitem [{\citenamefont {Gomes}\ \emph {et~al.}(2016)\citenamefont {Gomes},
  \citenamefont {\protect{De Silva}}, \citenamefont {Dutta}, \citenamefont
  {Clay},\ and\ \citenamefont {Mazumdar}}]{Gomes16a}%
  \BibitemOpen
  \bibfield  {author} {\bibinfo {author} {\bibfnamefont {N.}~\bibnamefont
  {Gomes}}, \bibinfo {author} {\bibfnamefont {W.~W.}\ \bibnamefont {\protect{De
  Silva}}}, \bibinfo {author} {\bibfnamefont {T.}~\bibnamefont {Dutta}},
  \bibinfo {author} {\bibfnamefont {R.~T.}\ \bibnamefont {Clay}},\ and\
  \bibinfo {author} {\bibfnamefont {S.}~\bibnamefont {Mazumdar}},\ }\bibfield
  {title} {\bibinfo {title} {Coulomb enhanced superconducting pair correlations
  in the frustrated quarter-filled band},\ }\href@noop {} {\bibfield  {journal}
  {\bibinfo  {journal} {Phys.\ Rev.\ B}\ }\textbf {\bibinfo {volume} {93}},\
  \bibinfo {pages} {165110} (\bibinfo {year} {2016})}\BibitemShut {NoStop}%
\bibitem [{\citenamefont {Venderley}\ and\ \citenamefont
  {Kim}(2019)}]{Venderley19a}%
  \BibitemOpen
  \bibfield  {author} {\bibinfo {author} {\bibfnamefont {J.}~\bibnamefont
  {Venderley}}\ and\ \bibinfo {author} {\bibfnamefont {E.-A.}\ \bibnamefont
  {Kim}},\ }\bibfield  {title} {\bibinfo {title} {Density matrix
  renormalization group study of superconductivity in the triangular lattice
  hubbard model},\ }\href@noop {} {\bibfield  {journal} {\bibinfo  {journal}
  {Phys.\ Rev.\ B}\ }\textbf {\bibinfo {volume} {100}},\ \bibinfo {pages}
  {060506(R)} (\bibinfo {year} {2019})}\BibitemShut {NoStop}%
\bibitem [{\citenamefont {Luther}\ and\ \citenamefont
  {Emery}(1974)}]{Luther74a}%
  \BibitemOpen
  \bibfield  {author} {\bibinfo {author} {\bibfnamefont {A.}~\bibnamefont
  {Luther}}\ and\ \bibinfo {author} {\bibfnamefont {V.~J.}\ \bibnamefont
  {Emery}},\ }\bibfield  {title} {\bibinfo {title} {Backward scattering in the
  one-dimensional electron gas},\ }\href@noop {} {\bibfield  {journal}
  {\bibinfo  {journal} {Phys.\ Rev.\ Lett.}\ }\textbf {\bibinfo {volume}
  {33}},\ \bibinfo {pages} {589} (\bibinfo {year} {1974})}\BibitemShut
  {NoStop}%
\bibitem [{\citenamefont {Noack}\ \emph {et~al.}(1994)\citenamefont {Noack},
  \citenamefont {White},\ and\ \citenamefont {Scalapino}}]{Noack94a}%
  \BibitemOpen
  \bibfield  {author} {\bibinfo {author} {\bibfnamefont {R.~M.}\ \bibnamefont
  {Noack}}, \bibinfo {author} {\bibfnamefont {S.~R.}\ \bibnamefont {White}},\
  and\ \bibinfo {author} {\bibfnamefont {D.~J.}\ \bibnamefont {Scalapino}},\
  }\bibfield  {title} {\bibinfo {title} {Correlations in a two-chain {H}ubbard
  model},\ }\href@noop {} {\bibfield  {journal} {\bibinfo  {journal} {Phys.\
  Rev.\ Lett.}\ }\textbf {\bibinfo {volume} {73}},\ \bibinfo {pages} {882}
  (\bibinfo {year} {1994})}\BibitemShut {NoStop}%
\bibitem [{\citenamefont {Noack}\ \emph {et~al.}(1997)\citenamefont {Noack},
  \citenamefont {Bulut}, \citenamefont {Scalapino},\ and\ \citenamefont
  {Zacher}}]{Noack97a}%
  \BibitemOpen
  \bibfield  {author} {\bibinfo {author} {\bibfnamefont {R.~M.}\ \bibnamefont
  {Noack}}, \bibinfo {author} {\bibfnamefont {N.}~\bibnamefont {Bulut}},
  \bibinfo {author} {\bibfnamefont {D.~J.}\ \bibnamefont {Scalapino}},\ and\
  \bibinfo {author} {\bibfnamefont {M.~G.}\ \bibnamefont {Zacher}},\ }\bibfield
   {title} {\bibinfo {title} {Enhanced \protect{$d_{x^2-y^2}$} pairing
  correlations in the two-leg {H}ubbard ladder},\ }\href@noop {} {\bibfield
  {journal} {\bibinfo  {journal} {Phys.\ Rev.\ B}\ }\textbf {\bibinfo {volume}
  {56}},\ \bibinfo {pages} {7162} (\bibinfo {year} {1997})}\BibitemShut
  {NoStop}%
\bibitem [{\citenamefont {Balents}\ and\ \citenamefont
  {Fisher}(1996)}]{Balents96a}%
  \BibitemOpen
  \bibfield  {author} {\bibinfo {author} {\bibfnamefont {L.}~\bibnamefont
  {Balents}}\ and\ \bibinfo {author} {\bibfnamefont {M.~P.~A.}\ \bibnamefont
  {Fisher}},\ }\bibfield  {title} {\bibinfo {title} {Weak-coupling phase
  diagram of the two-chain {H}ubbard model},\ }\href@noop {} {\bibfield
  {journal} {\bibinfo  {journal} {Phys.\ Rev.\ B}\ }\textbf {\bibinfo {volume}
  {53}},\ \bibinfo {pages} {12133} (\bibinfo {year} {1996})}\BibitemShut
  {NoStop}%
\bibitem [{\citenamefont {\protect{Le Hur}}\ and\ \citenamefont
  {Rice}(2009)}]{Hur09a}%
  \BibitemOpen
  \bibfield  {author} {\bibinfo {author} {\bibfnamefont {K.}~\bibnamefont
  {\protect{Le Hur}}}\ and\ \bibinfo {author} {\bibfnamefont {T.~M.}\
  \bibnamefont {Rice}},\ }\bibfield  {title} {\bibinfo {title}
  {Superconductivity close to the {M}ott state: From condensed-matter systems
  to the superfluidity in optical lattices},\ }\href@noop {} {\bibfield
  {journal} {\bibinfo  {journal} {Ann. Phys.}\ }\textbf {\bibinfo {volume}
  {324}},\ \bibinfo {pages} {1452} (\bibinfo {year} {2009})}\BibitemShut
  {NoStop}%
\bibitem [{\citenamefont {Dolfi}\ \emph {et~al.}(2015)\citenamefont {Dolfi},
  \citenamefont {Bauer}, \citenamefont {Keller},\ and\ \citenamefont
  {Troyer}}]{Dolfi15a}%
  \BibitemOpen
  \bibfield  {author} {\bibinfo {author} {\bibfnamefont {M.}~\bibnamefont
  {Dolfi}}, \bibinfo {author} {\bibfnamefont {B.}~\bibnamefont {Bauer}},
  \bibinfo {author} {\bibfnamefont {S.}~\bibnamefont {Keller}},\ and\ \bibinfo
  {author} {\bibfnamefont {M.}~\bibnamefont {Troyer}},\ }\bibfield  {title}
  {\bibinfo {title} {Pair correlations in doped {H}ubbard ladders},\
  }\href@noop {} {\bibfield  {journal} {\bibinfo  {journal} {Phys.\ Rev.\ B}\
  }\textbf {\bibinfo {volume} {92}},\ \bibinfo {pages} {195139} (\bibinfo
  {year} {2015})}\BibitemShut {NoStop}%
\bibitem [{\citenamefont {Gannot}\ \emph {et~al.}(2020)\citenamefont {Gannot},
  \citenamefont {Jiang},\ and\ \citenamefont {Kivelson}}]{Gannot20a}%
  \BibitemOpen
  \bibfield  {author} {\bibinfo {author} {\bibfnamefont {Y.}~\bibnamefont
  {Gannot}}, \bibinfo {author} {\bibfnamefont {Y.-F.}\ \bibnamefont {Jiang}},\
  and\ \bibinfo {author} {\bibfnamefont {S.~A.}\ \bibnamefont {Kivelson}},\
  }\bibfield  {title} {\bibinfo {title} {Hubbard ladders at small \protect{U}
  revisited},\ }\href@noop {} {\bibfield  {journal} {\bibinfo  {journal}
  {Phys.\ Rev.\ B}\ }\textbf {\bibinfo {volume} {102}},\ \bibinfo {pages}
  {115136} (\bibinfo {year} {2020})}\BibitemShut {NoStop}%
\bibitem [{\citenamefont {Emery}\ \emph {et~al.}(1997)\citenamefont {Emery},
  \citenamefont {Kivelson},\ and\ \citenamefont {Zachar}}]{Emery97a}%
  \BibitemOpen
  \bibfield  {author} {\bibinfo {author} {\bibfnamefont {V.~J.}\ \bibnamefont
  {Emery}}, \bibinfo {author} {\bibfnamefont {S.~A.}\ \bibnamefont
  {Kivelson}},\ and\ \bibinfo {author} {\bibfnamefont {O.}~\bibnamefont
  {Zachar}},\ }\bibfield  {title} {\bibinfo {title} {Spin-gap proximity effect
  mechanism of high-temperature superconductivity},\ }\href@noop {} {\bibfield
  {journal} {\bibinfo  {journal} {Phys.\ Rev.\ B}\ }\textbf {\bibinfo {volume}
  {56}},\ \bibinfo {pages} {6120} (\bibinfo {year} {1997})}\BibitemShut
  {NoStop}%
\bibitem [{\citenamefont {Feiguin}\ \emph {et~al.}(2008)\citenamefont
  {Feiguin}, \citenamefont {White}, \citenamefont {Scalapino},\ and\
  \citenamefont {Affleck}}]{Feiguin08a}%
  \BibitemOpen
  \bibfield  {author} {\bibinfo {author} {\bibfnamefont {A.~E.}\ \bibnamefont
  {Feiguin}}, \bibinfo {author} {\bibfnamefont {S.~R.}\ \bibnamefont {White}},
  \bibinfo {author} {\bibfnamefont {D.~J.}\ \bibnamefont {Scalapino}},\ and\
  \bibinfo {author} {\bibfnamefont {I.}~\bibnamefont {Affleck}},\ }\bibfield
  {title} {\bibinfo {title} {Pairing symmetry and {J}osephson current in doped
  2-leg \protect{$t$-$J$} ladders},\ }\href@noop {} {\bibfield  {journal}
  {\bibinfo  {journal} {Phys.\ Rev.\ Lett.}\ }\textbf {\bibinfo {volume}
  {101}},\ \bibinfo {pages} {217001} (\bibinfo {year} {2008})}\BibitemShut
  {NoStop}%
\bibitem [{\citenamefont {Dagotto}\ and\ \citenamefont
  {Rice}(1996)}]{Dagotto96a}%
  \BibitemOpen
  \bibfield  {author} {\bibinfo {author} {\bibfnamefont {E.}~\bibnamefont
  {Dagotto}}\ and\ \bibinfo {author} {\bibfnamefont {T.~M.}\ \bibnamefont
  {Rice}},\ }\bibfield  {title} {\bibinfo {title} {Surprises on the way from
  one- to two-dimensional quantum magnets: the ladder materials},\ }\href@noop
  {} {\bibfield  {journal} {\bibinfo  {journal} {Science}\ }\textbf {\bibinfo
  {volume} {271}},\ \bibinfo {pages} {618} (\bibinfo {year}
  {1996})}\BibitemShut {NoStop}%
\bibitem [{\citenamefont {Jeckelmann}\ \emph {et~al.}(1998)\citenamefont
  {Jeckelmann}, \citenamefont {Scalapino},\ and\ \citenamefont
  {White}}]{Jeckelmann98a}%
  \BibitemOpen
  \bibfield  {author} {\bibinfo {author} {\bibfnamefont {E.}~\bibnamefont
  {Jeckelmann}}, \bibinfo {author} {\bibfnamefont {D.~J.}\ \bibnamefont
  {Scalapino}},\ and\ \bibinfo {author} {\bibfnamefont {S.~R.}\ \bibnamefont
  {White}},\ }\bibfield  {title} {\bibinfo {title} {Comparison of different
  ladder models},\ }\href@noop {} {\bibfield  {journal} {\bibinfo  {journal}
  {Phys.\ Rev.\ B}\ }\textbf {\bibinfo {volume} {58}},\ \bibinfo {pages} {9492}
  (\bibinfo {year} {1998})}\BibitemShut {NoStop}%
\bibitem [{\citenamefont {Nishimoto}\ \emph {et~al.}(2002)\citenamefont
  {Nishimoto}, \citenamefont {Jeckelmann},\ and\ \citenamefont
  {Scalapino}}]{Nishimoto02a}%
  \BibitemOpen
  \bibfield  {author} {\bibinfo {author} {\bibfnamefont {S.}~\bibnamefont
  {Nishimoto}}, \bibinfo {author} {\bibfnamefont {E.}~\bibnamefont
  {Jeckelmann}},\ and\ \bibinfo {author} {\bibfnamefont {D.~J.}\ \bibnamefont
  {Scalapino}},\ }\bibfield  {title} {\bibinfo {title} {Differences between
  hole and electron doping of a two-leg {C}u{O} ladder},\ }\href@noop {}
  {\bibfield  {journal} {\bibinfo  {journal} {Phys.\ Rev.\ B}\ }\textbf
  {\bibinfo {volume} {66}},\ \bibinfo {pages} {245109} (\bibinfo {year}
  {2002})}\BibitemShut {NoStop}%
\bibitem [{\citenamefont {Hirayama}\ \emph {et~al.}(2018)\citenamefont
  {Hirayama}, \citenamefont {Yamaji}, \citenamefont {Misawa},\ and\
  \citenamefont {Imada}}]{Hirayama18a}%
  \BibitemOpen
  \bibfield  {author} {\bibinfo {author} {\bibfnamefont {M.}~\bibnamefont
  {Hirayama}}, \bibinfo {author} {\bibfnamefont {Y.}~\bibnamefont {Yamaji}},
  \bibinfo {author} {\bibfnamefont {T.}~\bibnamefont {Misawa}},\ and\ \bibinfo
  {author} {\bibfnamefont {M.}~\bibnamefont {Imada}},\ }\bibfield  {title}
  {\bibinfo {title} {Ab initio effective {H}amiltonians for cuprate
  superconductors},\ }\href@noop {} {\bibfield  {journal} {\bibinfo  {journal}
  {Phys.\ Rev.\ B}\ }\textbf {\bibinfo {volume} {98}},\ \bibinfo {pages}
  {134501} (\bibinfo {year} {2018})}\BibitemShut {NoStop}%
\bibitem [{\citenamefont {Fishman}\ \emph {et~al.}(2020)\citenamefont
  {Fishman}, \citenamefont {White},\ and\ \citenamefont
  {Stoudenmire}}]{itensor}%
  \BibitemOpen
  \bibfield  {author} {\bibinfo {author} {\bibfnamefont {M.}~\bibnamefont
  {Fishman}}, \bibinfo {author} {\bibfnamefont {S.~R.}\ \bibnamefont {White}},\
  and\ \bibinfo {author} {\bibfnamefont {E.~M.}\ \bibnamefont {Stoudenmire}},\
  }\bibfield  {title} {\bibinfo {title} {The {I}{T}ensor software library for
  tensor network calculations}} (\bibinfo {year} {2020}),\ \bibinfo {note}
  {https://arxiv.org/abs/2007.14822}\BibitemShut {NoStop}%
\bibitem [{\citenamefont {Stoudenmire}\ and\ \citenamefont
  {White}(2013)}]{Stoudenmire13a}%
  \BibitemOpen
  \bibfield  {author} {\bibinfo {author} {\bibfnamefont {E.~M.}\ \bibnamefont
  {Stoudenmire}}\ and\ \bibinfo {author} {\bibfnamefont {S.~R.}\ \bibnamefont
  {White}},\ }\bibfield  {title} {\bibinfo {title} {Real-space parallel density
  matrix renormalization group},\ }\href@noop {} {\bibfield  {journal}
  {\bibinfo  {journal} {Phys.\ Rev.\ B}\ }\textbf {\bibinfo {volume} {87}},\
  \bibinfo {pages} {155137} (\bibinfo {year} {2013})}\BibitemShut {NoStop}%
\bibitem [{Sup()}]{Supplemental}%
  \BibitemOpen
  \href@noop {} {}\bibinfo {note} {See Supplemental Material at
  http://link.aps.org/supplemental/xx.xxxx for further details of
  calculations.}\BibitemShut {Stop}%
\bibitem [{\citenamefont {Eskes}\ and\ \citenamefont
  {Jefferson}(1993)}]{Eskes93a}%
  \BibitemOpen
  \bibfield  {author} {\bibinfo {author} {\bibfnamefont {H.}~\bibnamefont
  {Eskes}}\ and\ \bibinfo {author} {\bibfnamefont {J.~H.}\ \bibnamefont
  {Jefferson}},\ }\bibfield  {title} {\bibinfo {title} {Superexchange in the
  cuprates},\ }\href@noop {} {\bibfield  {journal} {\bibinfo  {journal} {Phys.\
  Rev.\ B}\ }\textbf {\bibinfo {volume} {48}},\ \bibinfo {pages} {9788}
  (\bibinfo {year} {1993})}\BibitemShut {NoStop}%
\bibitem [{\citenamefont {Noack}\ \emph {et~al.}(1996)\citenamefont {Noack},
  \citenamefont {White},\ and\ \citenamefont {Scalapino}}]{Noack96a}%
  \BibitemOpen
  \bibfield  {author} {\bibinfo {author} {\bibfnamefont {R.~M.}\ \bibnamefont
  {Noack}}, \bibinfo {author} {\bibfnamefont {S.~R.}\ \bibnamefont {White}},\
  and\ \bibinfo {author} {\bibfnamefont {D.~J.}\ \bibnamefont {Scalapino}},\
  }\bibfield  {title} {\bibinfo {title} {The ground state of the two-leg
  {H}ubbard ladder {A} density-matrix renormalization group study},\
  }\href@noop {} {\bibfield  {journal} {\bibinfo  {journal} {Physica C}\
  }\textbf {\bibinfo {volume} {270}},\ \bibinfo {pages} {281} (\bibinfo {year}
  {1996})}\BibitemShut {NoStop}%
\bibitem [{\citenamefont {White}\ \emph {et~al.}(2002)\citenamefont {White},
  \citenamefont {Affleck},\ and\ \citenamefont {Scalapino}}]{White02a}%
  \BibitemOpen
  \bibfield  {author} {\bibinfo {author} {\bibfnamefont {S.~R.}\ \bibnamefont
  {White}}, \bibinfo {author} {\bibfnamefont {I.}~\bibnamefont {Affleck}},\
  and\ \bibinfo {author} {\bibfnamefont {D.~J.}\ \bibnamefont {Scalapino}},\
  }\bibfield  {title} {\bibinfo {title} {Friedel oscillations and charge
  density waves in chains and ladders},\ }\href@noop {} {\bibfield  {journal}
  {\bibinfo  {journal} {Phys.\ Rev.\ B}\ }\textbf {\bibinfo {volume} {65}},\
  \bibinfo {pages} {165122} (\bibinfo {year} {2002})}\BibitemShut {NoStop}%
\bibitem [{\citenamefont {Hiroi}\ and\ \citenamefont
  {Takano}(1995)}]{Hiroi95a}%
  \BibitemOpen
  \bibfield  {author} {\bibinfo {author} {\bibfnamefont {Z.}~\bibnamefont
  {Hiroi}}\ and\ \bibinfo {author} {\bibfnamefont {M.}~\bibnamefont {Takano}},\
  }\bibfield  {title} {\bibinfo {title} {Absence of superconductivity in the
  doped antiferromagnetic spin-ladder compound \protect{(La,Sr)CuO$_{2.5}$}},\
  }\href@noop {} {\bibfield  {journal} {\bibinfo  {journal} {Nature}\ }\textbf
  {\bibinfo {volume} {377}},\ \bibinfo {pages} {41} (\bibinfo {year}
  {1995})}\BibitemShut {NoStop}%
\bibitem [{\citenamefont {Uehara}\ \emph {et~al.}(1996)\citenamefont {Uehara},
  \citenamefont {Nagata}, \citenamefont {Akimitsu}, \citenamefont {Takahashi},
  \citenamefont {Mori},\ and\ \citenamefont {Kinoshita}}]{Uehara96a}%
  \BibitemOpen
  \bibfield  {author} {\bibinfo {author} {\bibfnamefont {M.}~\bibnamefont
  {Uehara}}, \bibinfo {author} {\bibfnamefont {T.}~\bibnamefont {Nagata}},
  \bibinfo {author} {\bibfnamefont {J.}~\bibnamefont {Akimitsu}}, \bibinfo
  {author} {\bibfnamefont {H.}~\bibnamefont {Takahashi}}, \bibinfo {author}
  {\bibfnamefont {N.}~\bibnamefont {Mori}},\ and\ \bibinfo {author}
  {\bibfnamefont {K.}~\bibnamefont {Kinoshita}},\ }\bibfield  {title} {\bibinfo
  {title} {Superconductivity in the ladder material
  \protect{Sr$_{0.4}$Ca$_{13.6}$Cu$_{24}$O$_{41.84}$}},\ }\href@noop {}
  {\bibfield  {journal} {\bibinfo  {journal} {J. Phys. Soc. Jpn.}\ }\textbf
  {\bibinfo {volume} {65}},\ \bibinfo {pages} {2764} (\bibinfo {year}
  {1996})}\BibitemShut {NoStop}%
\bibitem [{\citenamefont {Nagata}\ \emph {et~al.}(1998)\citenamefont {Nagata},
  \citenamefont {Uehara}, \citenamefont {Goto}, \citenamefont {Akimitsu},
  \citenamefont {Motoyama}, \citenamefont {Eisaki}, \citenamefont {Uchida},
  \citenamefont {Takahashi}, \citenamefont {Nakanishi},\ and\ \citenamefont
  {Mori}}]{Nagata98a}%
  \BibitemOpen
  \bibfield  {author} {\bibinfo {author} {\bibfnamefont {T.}~\bibnamefont
  {Nagata}}, \bibinfo {author} {\bibfnamefont {M.}~\bibnamefont {Uehara}},
  \bibinfo {author} {\bibfnamefont {J.}~\bibnamefont {Goto}}, \bibinfo {author}
  {\bibfnamefont {J.}~\bibnamefont {Akimitsu}}, \bibinfo {author}
  {\bibfnamefont {N.}~\bibnamefont {Motoyama}}, \bibinfo {author}
  {\bibfnamefont {H.}~\bibnamefont {Eisaki}}, \bibinfo {author} {\bibfnamefont
  {S.}~\bibnamefont {Uchida}}, \bibinfo {author} {\bibfnamefont
  {H.}~\bibnamefont {Takahashi}}, \bibinfo {author} {\bibfnamefont
  {T.}~\bibnamefont {Nakanishi}},\ and\ \bibinfo {author} {\bibfnamefont
  {N.}~\bibnamefont {Mori}},\ }\bibfield  {title} {\bibinfo {title}
  {Pressure-induced dimensional crossover and superconductivity in the
  hole-doped two-leg ladder compound
  \protect{Sr$_{14-x}$Ca$_x$Cu$_{24}$O$_{41}$}},\ }\href@noop {} {\bibfield
  {journal} {\bibinfo  {journal} {Phys.\ Rev.\ Lett.}\ }\textbf {\bibinfo
  {volume} {81}},\ \bibinfo {pages} {1090} (\bibinfo {year}
  {1998})}\BibitemShut {NoStop}%
\bibitem [{\citenamefont {\protect{Vuleti\'c}}\ \emph
  {et~al.}(2006)\citenamefont {\protect{Vuleti\'c}}, \citenamefont
  {\protect{Korin-Hamzi\'c}}, \citenamefont {Ivek}, \citenamefont
  {\protect{Tomi\'c}}, \citenamefont {Dressel},\ and\ \citenamefont
  {Akimitsu}}]{Vuletic06a}%
  \BibitemOpen
  \bibfield  {author} {\bibinfo {author} {\bibfnamefont {T.}~\bibnamefont
  {\protect{Vuleti\'c}}}, \bibinfo {author} {\bibfnamefont {B.}~\bibnamefont
  {\protect{Korin-Hamzi\'c}}}, \bibinfo {author} {\bibfnamefont
  {T.}~\bibnamefont {Ivek}}, \bibinfo {author} {\bibfnamefont {S.}~\bibnamefont
  {\protect{Tomi\'c}}}, \bibinfo {author} {\bibfnamefont {B.~G.~M.}\
  \bibnamefont {Dressel}},\ and\ \bibinfo {author} {\bibfnamefont
  {J.}~\bibnamefont {Akimitsu}},\ }\bibfield  {title} {\bibinfo {title} {The
  spin-ladder and spin-chain system
  \protect{(La,Y,Sr,Ca)$_{14}$Cu$_{24}$O$_{41}$}: Electronic phases, charge and
  spin dynamics},\ }\href@noop {} {\bibfield  {journal} {\bibinfo  {journal}
  {Phys. Rep.}\ }\textbf {\bibinfo {volume} {428}},\ \bibinfo {pages} {169}
  (\bibinfo {year} {2006})}\BibitemShut {NoStop}%
\bibitem [{\citenamefont {Bugnet}\ \emph {et~al.}(2016)\citenamefont {Bugnet},
  \citenamefont {Loeffler}, \citenamefont {Hawthorn}, \citenamefont
  {Dabkowska}, \citenamefont {Luke}, \citenamefont {Schattschneider},
  \citenamefont {Sawatzky}, \citenamefont {Radtke},\ and\ \citenamefont
  {Botton}}]{Bugnet16a}%
  \BibitemOpen
  \bibfield  {author} {\bibinfo {author} {\bibfnamefont {M.}~\bibnamefont
  {Bugnet}}, \bibinfo {author} {\bibfnamefont {S.}~\bibnamefont {Loeffler}},
  \bibinfo {author} {\bibfnamefont {D.}~\bibnamefont {Hawthorn}}, \bibinfo
  {author} {\bibfnamefont {H.~A.}\ \bibnamefont {Dabkowska}}, \bibinfo {author}
  {\bibfnamefont {G.~M.}\ \bibnamefont {Luke}}, \bibinfo {author}
  {\bibfnamefont {P.}~\bibnamefont {Schattschneider}}, \bibinfo {author}
  {\bibfnamefont {G.~A.}\ \bibnamefont {Sawatzky}}, \bibinfo {author}
  {\bibfnamefont {G.}~\bibnamefont {Radtke}},\ and\ \bibinfo {author}
  {\bibfnamefont {G.~A.}\ \bibnamefont {Botton}},\ }\bibfield  {title}
  {\bibinfo {title} {Real-space localization and quantification of hole
  distribution in chain-ladder \protect{Sr$_3$Ca$_{11}$Cu$_{24}$O$_{41}$}
  superconductor},\ }\href@noop {} {\bibfield  {journal} {\bibinfo  {journal}
  {Sci. Adv.}\ }\textbf {\bibinfo {volume} {2}},\ \bibinfo {pages} {e1501652}
  (\bibinfo {year} {2016})}\BibitemShut {NoStop}%
\bibitem [{\citenamefont {Piskunov}\ \emph {et~al.}(2004)\citenamefont
  {Piskunov}, \citenamefont {J{\'e}rome}, \citenamefont {Auban-Senzier},
  \citenamefont {Wzietek},\ and\ \citenamefont {Yakubovsky}}]{Piskunov05a}%
  \BibitemOpen
  \bibfield  {author} {\bibinfo {author} {\bibfnamefont {Y.}~\bibnamefont
  {Piskunov}}, \bibinfo {author} {\bibfnamefont {D.}~\bibnamefont
  {J{\'e}rome}}, \bibinfo {author} {\bibfnamefont {P.}~\bibnamefont
  {Auban-Senzier}}, \bibinfo {author} {\bibfnamefont {P.}~\bibnamefont
  {Wzietek}},\ and\ \bibinfo {author} {\bibfnamefont {A.}~\bibnamefont
  {Yakubovsky}},\ }\bibfield  {title} {\bibinfo {title} {Spin excitations in
  the \protect{(Sr,Ca)$_{14}$Cu$_{24}$O$_{41}$} family of spin ladders:
  \protect{$^{63}$Cu} and \protect{$^{17}$O} \protect{NMR} studies under
  pressure},\ }\href@noop {} {\bibfield  {journal} {\bibinfo  {journal} {Phys.\
  Rev.\ B}\ }\textbf {\bibinfo {volume} {69}},\ \bibinfo {pages} {014510}
  (\bibinfo {year} {2004})}\BibitemShut {NoStop}%
\bibitem [{\citenamefont {Fujiwara}\ \emph {et~al.}(2009)\citenamefont
  {Fujiwara}, \citenamefont {Fujimaki}, \citenamefont {Uchida}, \citenamefont
  {Matsubayashi}, \citenamefont {Matsumoto},\ and\ \citenamefont
  {Uwatoko}}]{Fujiwara09a}%
  \BibitemOpen
  \bibfield  {author} {\bibinfo {author} {\bibfnamefont {N.}~\bibnamefont
  {Fujiwara}}, \bibinfo {author} {\bibfnamefont {Y.}~\bibnamefont {Fujimaki}},
  \bibinfo {author} {\bibfnamefont {S.}~\bibnamefont {Uchida}}, \bibinfo
  {author} {\bibfnamefont {K.}~\bibnamefont {Matsubayashi}}, \bibinfo {author}
  {\bibfnamefont {T.}~\bibnamefont {Matsumoto}},\ and\ \bibinfo {author}
  {\bibfnamefont {Y.}~\bibnamefont {Uwatoko}},\ }\bibfield  {title} {\bibinfo
  {title} {\protect{NMR} and \protect{NQR} study of pressure-induced
  superconductivity and the origin of critical-temperature enhancement in the
  spin-ladder cuprate \protect{Sr$_2$Ca$_{12}$Cu$_{24}$O$_{41}$}},\ }\href@noop
  {} {\bibfield  {journal} {\bibinfo  {journal} {Phys.\ Rev.\ B}\ }\textbf
  {\bibinfo {volume} {80}},\ \bibinfo {pages} {100503(R)} (\bibinfo {year}
  {2009})}\BibitemShut {NoStop}%
\bibitem [{\citenamefont {Arrigoni}\ \emph {et~al.}(2004)\citenamefont
  {Arrigoni}, \citenamefont {Fradkin},\ and\ \citenamefont
  {Kivelson}}]{Arrigoni04a}%
  \BibitemOpen
  \bibfield  {author} {\bibinfo {author} {\bibfnamefont {E.}~\bibnamefont
  {Arrigoni}}, \bibinfo {author} {\bibfnamefont {E.}~\bibnamefont {Fradkin}},\
  and\ \bibinfo {author} {\bibfnamefont {S.~A.}\ \bibnamefont {Kivelson}},\
  }\bibfield  {title} {\bibinfo {title} {Mechanism of high-temperature
  superconductivity in a striped {H}ubbard model},\ }\href@noop {} {\bibfield
  {journal} {\bibinfo  {journal} {Phys.\ Rev.\ B}\ }\textbf {\bibinfo {volume}
  {69}},\ \bibinfo {pages} {214519} (\bibinfo {year} {2004})}\BibitemShut
  {NoStop}%
\bibitem [{\citenamefont {Peets}\ \emph {et~al.}(2009)\citenamefont {Peets},
  \citenamefont {Hawthorn}, \citenamefont {Shen}, \citenamefont {Kim},
  \citenamefont {Ellis}, \citenamefont {Zhang}, \citenamefont {Komiya},
  \citenamefont {Ando}, \citenamefont {Sawatzky}, \citenamefont {Liang},
  \citenamefont {Bonn},\ and\ \citenamefont {Hardy}}]{Peets09a}%
  \BibitemOpen
  \bibfield  {author} {\bibinfo {author} {\bibfnamefont {D.~C.}\ \bibnamefont
  {Peets}}, \bibinfo {author} {\bibfnamefont {D.~G.}\ \bibnamefont {Hawthorn}},
  \bibinfo {author} {\bibfnamefont {K.~M.}\ \bibnamefont {Shen}}, \bibinfo
  {author} {\bibfnamefont {Y.-J.}\ \bibnamefont {Kim}}, \bibinfo {author}
  {\bibfnamefont {D.~S.}\ \bibnamefont {Ellis}}, \bibinfo {author}
  {\bibfnamefont {H.}~\bibnamefont {Zhang}}, \bibinfo {author} {\bibfnamefont
  {S.}~\bibnamefont {Komiya}}, \bibinfo {author} {\bibfnamefont
  {Y.}~\bibnamefont {Ando}}, \bibinfo {author} {\bibfnamefont {G.~A.}\
  \bibnamefont {Sawatzky}}, \bibinfo {author} {\bibfnamefont {R.}~\bibnamefont
  {Liang}}, \bibinfo {author} {\bibfnamefont {D.~A.}\ \bibnamefont {Bonn}},\
  and\ \bibinfo {author} {\bibfnamefont {W.~N.}\ \bibnamefont {Hardy}},\
  }\bibfield  {title} {\bibinfo {title} {X-ray absorption spectra reveal the
  inapplicability of the single-band {H}ubbard model to overdoped cuprate
  superconductors},\ }\href@noop {} {\bibfield  {journal} {\bibinfo  {journal}
  {Phys.\ Rev.\ Lett.}\ }\textbf {\bibinfo {volume} {103}},\ \bibinfo {pages}
  {087402} (\bibinfo {year} {2009})}\BibitemShut {NoStop}%
\bibitem [{\citenamefont {Sunko}(2009)}]{Sunko09a}%
  \BibitemOpen
  \bibfield  {author} {\bibinfo {author} {\bibfnamefont {D.~K.}\ \bibnamefont
  {Sunko}},\ }\bibfield  {title} {\bibinfo {title} {Destabilization of the
  {Z}hang-{R}ice singlet at optimal doping},\ }\href@noop {} {\bibfield
  {journal} {\bibinfo  {journal} {J. Exp. and Theor. Phys.}\ }\textbf {\bibinfo
  {volume} {109}},\ \bibinfo {pages} {652} (\bibinfo {year}
  {2009})}\BibitemShut {NoStop}%
\bibitem [{\citenamefont {Adolphs}\ \emph {et~al.}(2016)\citenamefont
  {Adolphs}, \citenamefont {Moser}, \citenamefont {Sawatzky},\ and\
  \citenamefont {Berciu}}]{Adolphs16a}%
  \BibitemOpen
  \bibfield  {author} {\bibinfo {author} {\bibfnamefont {C.~P.~J.}\
  \bibnamefont {Adolphs}}, \bibinfo {author} {\bibfnamefont {S.}~\bibnamefont
  {Moser}}, \bibinfo {author} {\bibfnamefont {G.~A.}\ \bibnamefont
  {Sawatzky}},\ and\ \bibinfo {author} {\bibfnamefont {M.}~\bibnamefont
  {Berciu}},\ }\bibfield  {title} {\bibinfo {title} {Non-{Z}hang-{R}ice singlet
  character of the first ionization state of \protect{T-CuO}},\ }\href@noop {}
  {\bibfield  {journal} {\bibinfo  {journal} {Phys.\ Rev.\ Lett.}\ }\textbf
  {\bibinfo {volume} {116}},\ \bibinfo {pages} {087002} (\bibinfo {year}
  {2016})}\BibitemShut {NoStop}%
\bibitem [{\citenamefont {Guerrero}\ \emph {et~al.}(1998)\citenamefont
  {Guerrero}, \citenamefont {Gubernatis},\ and\ \citenamefont
  {Zhang}}]{Guerrero98a}%
  \BibitemOpen
  \bibfield  {author} {\bibinfo {author} {\bibfnamefont {M.}~\bibnamefont
  {Guerrero}}, \bibinfo {author} {\bibfnamefont {J.~E.}\ \bibnamefont
  {Gubernatis}},\ and\ \bibinfo {author} {\bibfnamefont {S.}~\bibnamefont
  {Zhang}},\ }\bibfield  {title} {\bibinfo {title} {Quantum {M}onte {C}arlo
  study of hole binding and pairing correlations in the three-band {H}ubbard
  model},\ }\href@noop {} {\bibfield  {journal} {\bibinfo  {journal} {Phys.\
  Rev.\ B}\ }\textbf {\bibinfo {volume} {57}},\ \bibinfo {pages} {11980}
  (\bibinfo {year} {1998})}\BibitemShut {NoStop}%
\bibitem [{\citenamefont {Towns}\ \emph {et~al.}(2014)\citenamefont {Towns},
  \citenamefont {Cockerill}, \citenamefont {Dahan}, \citenamefont {Foster},
  \citenamefont {Gaither}, \citenamefont {Grimshaw}, \citenamefont {Hazlewood},
  \citenamefont {Lathrop}, \citenamefont {Lifka}, \citenamefont {Peterson},
  \citenamefont {Roskies}, \citenamefont {Scott},\ and\ \citenamefont
  {Wilkins-Diehr}}]{xsede}%
  \BibitemOpen
  \bibfield  {author} {\bibinfo {author} {\bibfnamefont {J.}~\bibnamefont
  {Towns}}, \bibinfo {author} {\bibfnamefont {T.}~\bibnamefont {Cockerill}},
  \bibinfo {author} {\bibfnamefont {M.}~\bibnamefont {Dahan}}, \bibinfo
  {author} {\bibfnamefont {I.}~\bibnamefont {Foster}}, \bibinfo {author}
  {\bibfnamefont {K.}~\bibnamefont {Gaither}}, \bibinfo {author} {\bibfnamefont
  {A.}~\bibnamefont {Grimshaw}}, \bibinfo {author} {\bibfnamefont
  {V.}~\bibnamefont {Hazlewood}}, \bibinfo {author} {\bibfnamefont
  {S.}~\bibnamefont {Lathrop}}, \bibinfo {author} {\bibfnamefont
  {D.}~\bibnamefont {Lifka}}, \bibinfo {author} {\bibfnamefont {G.~D.}\
  \bibnamefont {Peterson}}, \bibinfo {author} {\bibfnamefont {R.}~\bibnamefont
  {Roskies}}, \bibinfo {author} {\bibfnamefont {J.~R.}\ \bibnamefont {Scott}},\
  and\ \bibinfo {author} {\bibfnamefont {N.}~\bibnamefont {Wilkins-Diehr}},\
  }\bibfield  {title} {\bibinfo {title} {\protect{XSEDE}: Accelerating
  scientific discovery},\ }\href@noop {} {\bibfield  {journal} {\bibinfo
  {journal} {Computing in Science \& Engineering}\ }\textbf {\bibinfo {volume}
  {16}},\ \bibinfo {pages} {62} (\bibinfo {year} {2014})}\BibitemShut {NoStop}%
\end{thebibliography}
\end{document}